%% file: main.tex
\documentclass[usenatbib]{mnras}
\usepackage{newtxtext,newtxmath}
\usepackage[T1]{fontenc}

\DeclareRobustCommand{\VAN}[3]{#2}
\let\VANthebibliography\thebibliography
\def\thebibliography{\DeclareRobustCommand{\VAN}[3]{##3}\VANthebibliography}

\usepackage{amsmath}
\usepackage{graphicx}
\usepackage[usenames,dvipsnames]{xcolor}
\usepackage{mathtools}
\usepackage{gensymb}
\usepackage{tabularx}
\usepackage{subfigure}
\usepackage{physics}
\usepackage{subfiles}

\newcommand{\dobib}{\bibliography{references}}
\newcommand{\Da}{Damk{\"o}hler }

\title{Radiative Mixing Layers: Insights from Turbulent Combustion}

\author[B. Tan, S.P. Oh and M. Gronke] {
  Brent Tan$^1$\thanks{E-mail: zunyibrent@physics.ucsb.edu},
  S. Peng Oh$^1$,
  and Max Gronke$^{1,2}$\thanks{Hubble fellow}\\
    $^1$University of California - Santa Barbara,
    Department of Physics, CA 93106-9530, USA\\
    $^2$Department of Physics \& Astronomy, Johns Hopkins University, 
    Bloomberg Center, 3400 N. Charles St., Baltimore, MD 21218, USA
}
\date{Accepted XXX. Received YYY; in original form ZZZ}
\pubyear{2020}

\begin{document}
\renewcommand{\dobib}{}
\label{firstpage}
\pagerange{\pageref{firstpage}--\pageref{lastpage}}
\maketitle

\begin{abstract}
Radiative mixing layers arise wherever multiphase gas, shear, and radiative cooling are present. Simulations show that in steady state, thermal advection from the hot phase balances radiative cooling. However, many features are puzzling. For instance, hot gas entrainment appears to be numerically converged despite the scale-free, fractal structure of such fronts being unresolved. Additionally, the hot gas heat flux has a characteristic velocity $v_{\rm in} \approx c_{\rm s,cold} (t_{\rm cool}/t_{\rm sc,cold})^{-1/4}$ whose strength and scaling are not intuitive. We revisit these issues in 1D and 3D hydrodynamic simulations. We find that over-cooling only happens if numerical diffusion dominates thermal transport; convergence is still possible even when the Field length is unresolved. A deeper physical understanding of radiative fronts can be obtained by exploiting parallels between mixing layers and turbulent combustion, which has well-developed theory and abundant experimental data. A key parameter is the \Da number ${\rm Da} = \tau_{\rm turb}/t_{\rm cool}$, the ratio of the outer eddy turnover time to the cooling time. Once ${\rm Da} > 1$, the front fragments into a multiphase medium. Just as for scalar mixing, the eddy turnover time sets the mixing rate, independent of small scale diffusion. For this reason, thermal conduction often has limited impact. We show that $v_{\rm in}$ and the effective emissivity can be understood in detail by adapting combustion theory scalings. Mean density and temperature profiles can also be reproduced remarkably well by mixing length theory. These results have implications for the structure and survival of cold gas in many settings, and resolution requirements for large scale galaxy simulations. 
\end{abstract}



\subfile{sections/1_introduction}

\subfile{sections/2_methods}

\subfile{sections/3_1D}

\subfile{sections/4_combustion}

\subfile{sections/5_3D}

\subfile{sections/6_discussion}

\subfile{sections/7_ackd}
\newpage
\bibliography{references}
\newpage

\subfile{sections/8_appendix}

\bsp
\label{lastpage}
\end{document}

%% file: sections/1_introduction.tex
\section{Introduction} \label{sect:intro}
Multiphase media are ubiquitous in astrophysics. Interfaces between different phases are not infinitely sharp, but thickened by energy transport processes such as thermal conduction \citep{borkowski90,gnat10} and collisionless cosmic ray (CR) scattering  \citep{wiener17-cold-clouds}. Shear flows further structure the interface, by driving the Kelvin-Helmholtz (KH) instability, which seeds turbulence and fluid mixing. In ideal hydrodynamics, the KH instability is scale free. However, non-ideal processes, such as viscosity, can set a characteristic scale. Perhaps the most important of these non-ideal processes is radiative cooling, which typically is very strong in mixed gas at temperatures intermediate between the two stable phases. For instance, under coronal conditions the cooling curve peaks at $T \sim 10^{5}\,$K, intermediate between the $T\sim 10^{4}\,$K and $T\sim 10^{6}\,$K phases. Radiative turbulent mixing layers (TMLs) then arise where the exchange of mass, momentum and energy between phases is governed by the interaction between turbulence and radiative cooling. This has many important physical and observational consequences. For example, in the circumgalactic medium (CGM), such physics governs the growth or destruction of cold clouds embedded in a hot wind \citep{klein94,mellema02,pittard05,cooper09,scannapieco15,schneider16, gronke18,gronke19}, and the survival of cold streams inflowing from cosmological accretion \citep{2018A&A...610A..75C,mandelker20a}. TMLs also `host' $T \sim 10^{5}\,$K gas which could explain the abundance of OVI seen in galaxy halos, despite the fact that it is thermally unstable \citep{slavin93}. In addition, TMLs play crucial roles in the ISM (e.g., in supernova explosions), galaxy clusters (e.g., in the interface between optical filaments and the intracluster medium), AGN environments (e.g., chaotic cold accretion on the AGN, \citealp{2013MNRAS.432.3401G}); survival and stability of AGN jets (\citealp{1997ApJ...483..121H}), and many other astrophysical settings. 

Despite their ubiquity and importance, radiative mixing layers have received relatively little attention compared to adiabatic simulations of the Kelvin Helmholtz instability. \citet{begelman90} wrote an early analytic paper suggesting that radiative mixing layers are characterized by a mean temperature $\bar{T} \sim (T_{\rm hot} T_{\rm cold})^{1/2}$ and width $l \sim v_{\rm t} t_{\rm cool}$, where $v_{\rm t}$ is the turbulent velocity and $t_{\rm cool}$ is the cooling time of the mixed gas. In a series of papers \citep{kwak10, kwak11, henley12, kwak15}, Kwak and collaborators ran 2D hydrodynamic simulations and compared to observed line column densities and ratios, but not to analytic theory. \citet{esquivel06} ran 3D MHD simulations, but not for long enough for effective mixing (or a stable equilibrium) to develop. 

More recently, \citet{ji19} performed 3D hydrodynamic and MHD simulations, including both photoionization and non-equilibrium ionization. Interestingly, they found strong discrepancies with analytic models even in the purely hydrodynamic regime, with characteristic inflow and turbulent velocities much less than the shear velocity, and of order the cold gas sound speed, $c_{\rm s,cold}$. They also found the layer width $l \propto t_{\rm cool}^{1/2}$ rather than $l \propto t_{\rm cool}$, implying a surface brightness and mass entrainment velocity $Q, v_{\rm in} \propto t_{\rm cool}^{-1/2}$. They also found that previous analytic scalings (e.g., that the column density is independent of density or metallicity) do not agree with simulations. Subsequently, \citet{gronke18,gronke19} looked at mass growth of cold clouds embedded in a wind and found similar inflow velocities $v_{\rm in} \sim c_{\rm s,cold}$, but with a different scaling $Q, v_{\rm in} \propto t_{\rm cool}^{-1/4}$, which has also been seen by \citet{mandelker20a,fielding20}.  \citet{fielding20} ran a suite of 3D hydrodynamic simulations similar to \citet{ji19}, and highlighted the fractal nature of the interface; they derived a formula for $v_{\rm in}$ based on this observation. 

The situation is far from resolved. In our opinion, some of the biggest outstanding questions are: 
\begin{itemize}
\item{{\it Scalings}. In previous work \citep{gronke19}, we found: 
\begin{equation}  
    v_{\rm in} \approx 0.2  c_{\rm s,cold} \left(\frac{t_{\rm cool}}{t_{\rm sc,cold}} \right)^{-1/4} = 0.2 c_{\rm s,cold} \left(\frac{c_{\rm s,cold} t_{\rm cool}}{L} \right)^{-1/4} ,
    \label{eq:vin} 
\end{equation}
where $L$ is a characteristic length scale, and $c_{\rm s,cold}$ is the  sound speed of the cold gas. These scalings are not intuitive, and do not contain the shear velocity $v_{\rm shear}$ and overdensity $\chi$  which  might be expected to play a role in the hot gas entrainment rate. What is their origin? And why are there discrepant scalings of $v_{\rm in} \propto t_{\rm cool}^{-1/2}$ \citep{ji19} and $v_{\rm in} \propto t_{\rm cool}^{-1/4}$ \citep{gronke18,gronke19,fielding20}?}  

\item{{\it Energetics.} In steady state, cooling in the mixing layer is balanced by enthalpy flux from the hot gas, at a rate $\sim 5/2 P v_{\rm in}$. 
To order of magnitude, inflow of the hot gas occurs at roughly the cold gas sound speed, $v_{\rm in} \sim c_{\rm s,cold}$, as seen in Eq.~\eqref{eq:vin}. This may seem surprising, since  it is far below the maximum rate $\sim c_{\rm s,hot}$ at which the hot gas can potentially deliver enthalpy. For instance, saturated thermal conduction has a heat flux $\sim P c_{\rm s,hot}$. Why is turbulent heat diffusion so inefficient? The simulations of \citep{gronke18,gronke19} suggested that pulsations of the cold gas cloud (driven out of pressure balance with surroundings by radiative cooling were responsible for drawing in hot gas, in which case $c_{\rm s,cold}$ might be a natural velocity scale. However, it is not clear why $v_{\rm in} \sim c_{\rm s,cold}$ should be similar in a plane parallel shear layer, where the velocity shear drives mixing.}

\item {\it Robustness to Resolution.} Perhaps the most surprising  feature  of  the simulations is the robustness of $v_{\rm in}$ (or equivalently, the surface brightness $Q$) to numerical resolution. Most cooling occurs in the thermal front, where the gas  transitions between the thermally  stable phases $T_{\rm cold}$ and $T_{\rm hot}$. It is widely accepted that for numerical convergence, such transition layers must have finite thickness (by explicit inclusion of thermal conduction) and moreover that these fronts must be numerically resolved by at least 4 cells \citep{koyama04}. Otherwise, cooling gas fragments to the grid scale, and the total amount of cooling is resolution dependent (`numerical overcooling'). 
Most simulations mentioned above do not include explicit  thermal conduction and most of the emission occurs in zones $\sim 1$ cell thick -- yet the surface brightness $Q$ appears numerically converged. Surprisingly, the value of $v_{\rm in}$ in the high resolution calculations of a single plane parallel mixing layer \citep{ji19,fielding20} agree with the results of \citet{gronke18,gronke19,gronke20}, which embed a macroscopic cloud in a wind. In the latter case, by necessity resolution is orders of magnitude worse and the entire mixing layer is essentially unresolved. Simulations of radiative cooling in a turbulent, thermally bistable medium also show convergence in global quantities such as the density PDF, despite no explicit thermal conduction and lack of convergence in cold gas morphology \citep{gazol05}. The morphology of the mixing region is a strong function of resolution. For instance, the area of the cooling surface increases with resolution, and recently  \citet{fielding20} demonstrated that the area is a fractal with $A \propto  \lambda^{-1/2}$, where  $\lambda$ is the smoothing scale. Since the volume of the cooling region scales as $\sim A \lambda \sim \lambda^{1/2}$, one would expect the total cooling rate to  be resolution dependent. Somehow  it is not, even when characteristic scales such as the cooling length $c_{\rm s} t_{\rm cool}$ are highly under-resolved. It is critical to understand this, particularly in the context of  prescribing resolution requirements for larger scale simulations of  galaxy formation. For instance, the circumgalactic medium (CGM) in present day state of the art galaxy simulations is unconverged, with HI column densities continually rising with resolution \citep{VandeVoort2018,hummels19,Peeples2018,nelson20,mandelker19b}.  

\end{itemize}

In this paper, we exploit the close parallels between a two-phase turbulent radiative front and a turbulent combustion front to understand the above issues. In the parlance of combustion fronts, hot gas is the `fuel' and cold gas is the `oxidizer' which `burn' to give `ashes' (more cold gas). There is an extensive literature on combustion which not only has theoretical and computational underpinnings, but vast experimental backing as well -- a critical component in a situation where it is unclear whether numerical hydrodynamic codes can attain the required dynamic range. We explore the distinction between laminar and turbulent radiative fronts, with a particular focus on numerical convergence and robustness to resolution. 

The structure of the paper is as follows. In \S\ref{sect:methods}, we detail the implementation of radiative cooling and thermal conduction in our simulations. In \S\ref{sect:1D}, we describe 1D simulations with radiative cooling and conduction which probe the dependence of laminar fronts to resolution. In \S\ref{sect:combustion}, we explore parallels between radiative fronts and turbulent combustion, and review findings from the turbulent combustion literature. Based on this, we also develop an analytic model of radiative TMLs. In \S\ref{sect:turb}, we show 3D simulations which develop a turbulent mixing layer with radiative cooling via the Kelvin-Helmholtz instability. We compare our results to analytic predictions, and a 1D mixing length model. Finally, we conclude in \S\ref{sect:disc}.

\dobib

%% file: sections/2_methods.tex
\section{Methods} \label{sect:methods}

We carry out our simulations using the publicly available MHD code Athena\verb!++! \citep{Stone2020}. All simulations are run on regular Cartesian grids and use the HLLC Riemann solver. The individual simulation setups of 1D laminar fronts and 3D turbulent fronts are described separately in \S\ref{sect:1Dsetup} and \S\ref{sect:3Dsetup}. Here, we describe how we implement radiative cooling, present in all our simulations, and thermal conduction, present in all 1D simulations and a subset of 3D simulations (\S\ref{sect:3dthermalconduction}).

\subsection{Radiative Cooling}

\begin{figure}
    \centering
    \includegraphics[width=\columnwidth]{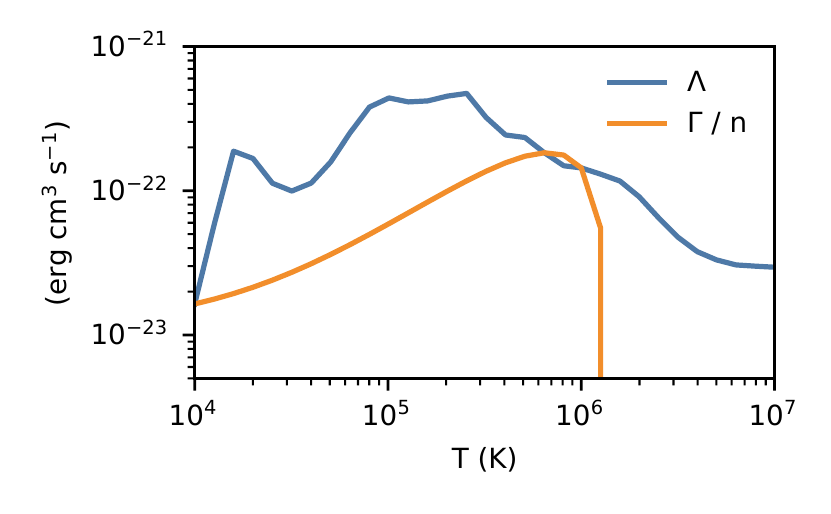}
    \caption{Heating and cooling rates as a function of temperature, with a temperature floor set at $10^4$ K. The two stable phases are at $10^4$ K and $10^6$ K.}
    \label{fig:coolcurve}
\end{figure}
The net cooling rate per unit volume is usually written as
\begin{equation}
    \rho \mathcal{L} = n^2 \Lambda - n\Gamma ,
\end{equation}
where $\Lambda$ is the cooling function and $\Gamma$ is the heating rate. For radiative cooling, we assume conditions of collisional ionization equilibrium and solar metallicity ($X=0.7$, $Z=0.02$). We obtain our cooling curve by performing a piece-wise power law fit to the cooling table given in \cite{Gnat_2007} over 40 logarithmically spaced temperature bins, starting from a temperature floor of $10^4$~K. We then implement the fast and robust exact cooling algorithm described in \citet{townsend}. We also add in a density dependent heating rate such that we have a thermally bistable medium. The cooling and heating curves that we used are shown in Fig.~\ref{fig:coolcurve}. While the inclusion of heating for a formally bistable medium is important in comparing to exact analytic solutions for the 1D front, it is inconsequential to the cooling rates in 3D simulations; we thus resort to a fixed temperature floor as well as setting the cooling rate in the hot medium ($T> 0.5 T_{\rm hot}$) to be zero in our 3D simulations. For some tests, we used a different shape of the cooling curve, which we specify in the relevant respective section.

\subsection{Thermal Conduction}

The conductive heat flux is $Q=-\kappa \grad T$, where the thermal conductivity of an ionized plasma is given by \cite{spitzer}: 
\begin{align}
    \kappa_{\rm sp} &= 5.7 \times 10^{-7} \; T^{2.5} \; \text{erg cm$^{-1}$ s$^{-1}$ K$^{-1}$} .
\end{align}

Instead of using the above temperature dependent conductivity, we assume a constant conductivity equivalent to the value of $\kappa_{\rm sp}$ at the temperature of the warm gas, $T= 0.8 \times 10^5$ K. This is numerically convenient but does not significantly change the results presented. The conductivity we use where applicable is hence
\begin{align} \label{eq:conduction} 
    \kappa &= 10^6 \; \text{erg cm$^{-1}$ s$^{-1}$ K$^{-1}$} .
\end{align}

As thermal conduction is a diffusive process, it is normally computationally expensive to implement. We employ a two moment approximation method for conduction similar to the approach used for implementing cosmic rays in \cite{jiang18}. This is done by introducing a second equation
\begin{equation}
  \frac{1}{V_{\rm m}^2} \frac{\partial Q}{\partial t} + \grad E = -\frac{\rho Q}{(\gamma -1)\kappa} ,
\end{equation}
with an effective propagation speed $V_{\rm m}$. The latter represents the ballistic velocity of free electrons, which is $\sim \sqrt{m_{p}/m_{e}} \sim 45$ times larger than the gas sound speed\footnote[1]{The analogous quantity in \citet{jiang18} is the reduced speed of light for free-streaming cosmic rays.}. In the limit that $V_{\rm m}$ goes to infinity, the equation reduces to the usual equation for heat conduction. As long as $V_{\rm m}$ is large compared to the speeds in the simulation, the solution is a good approximation to the true solution. We check that our results are converged with respect to $V_{\rm m}$ (cf. Appendix~\ref{appendix:conduction}). The timestep of this approach scales as $O(\Delta x)$, compared to traditional explicit schemes which scale as $O(\Delta x^2)$. Implicit schemes which also have a linear scaling with resolution are constrained by the fact that they require matrix inversion over the whole simulation domain, which can be slow and hinders parallelization. The module employs operator splitting to compute the transport and source terms, using a two step van-Leer time integrator; the  source term is  added implicitly. The algorithm is presented and discussed in Jiang et al. 2021 (in preparation); we thank Y.F. Jiang for providing the code in advance of publication. 

\subsection{Parameter studies}

In our simulations, we vary thermal conduction and cooling strength, and will henceforth refer to them as constant multiples of the fiducial values described above. To adjust the cooling strength, we change the normalization of the cooling curve via multiplication by a constant prefactor $\Lambda_0$. Physically, a change in the cooling time is usually due to a change in the ambient pressure; adjusting the normalization for the cooling curve achieves the same result and is more  numerically convenient. In the stratified CGM, the cooling time is a function of radius. Similarly, to adjust conduction, we multiply the conductivity by a prefactor $\kappa_0$. The cooling function and conductivity in a given simulation are thus given by
\begin{align}
    \Lambda(T) &= \Lambda_0 \Lambda_{\rm fid}(T) \\
    \kappa &= \kappa_0\kappa_{\rm fid} ,
\end{align}
where $\Lambda_{\rm fid}$ and $\kappa_{\rm fid}$ are the fiducial cooling profile and conductivity given in  Fig.~\ref{fig:coolcurve} and Eq.~\eqref{eq:conduction} respectively.

In radiative mixing layers, radiative cooling is balanced by enthalpy flux \citep{ji19}, 
\begin{equation} 
    Q \approx \frac{5}{2} P v_{\rm in} ,
\end{equation}
where $Q$ is the surface brightness. Hence, measuring $Q$ or $v_{\rm in}$ are equivalent. We focus on measuring $Q$ as it is a frame-independent quantity.
 
\dobib

%% file: sections/3_1D.tex
\section{1D Simulations: Laminar Fronts}  \label{sect:1D} 

A large focus of this paper is on resolution requirements and convergence issues. As we shall see, the structure of the front depends strongly on whether the flow is laminar or turbulent, and on the dominant heat diffusion mechanism: thermal conduction, turbulence, or numerical diffusion. We first study the behavior of laminar flows in 1D simulations with thermal conduction and cooling. In the parlance of turbulent combustion discussed at length in \S\ref{sect:combustion}, this gives us insight into the behavior of the laminar flame speed $S_{\rm L}$ and associated convergence issues. Conventional wisdom (e.g, \citealt{koyama04}) holds that it is necessary to (a) include explicit thermal conduction, and (b) resolve the {\it smallest} Field length in the  problem (usually of  the coldest gas), in order for calculations to be numerically converged. This is unequivocally true if we seek numerically converged temperature and density profiles.  However, we shall see that if we merely  seek numerical convergence in  the mass flux $j_{\rm x}$ and  hence the surface  brightness $Q$, there are some subtleties which relax this requirement. 

\subsection{Setup} \label{sect:1Dsetup} 

\begin{figure} 
    \centering
    \includegraphics[width=\columnwidth]{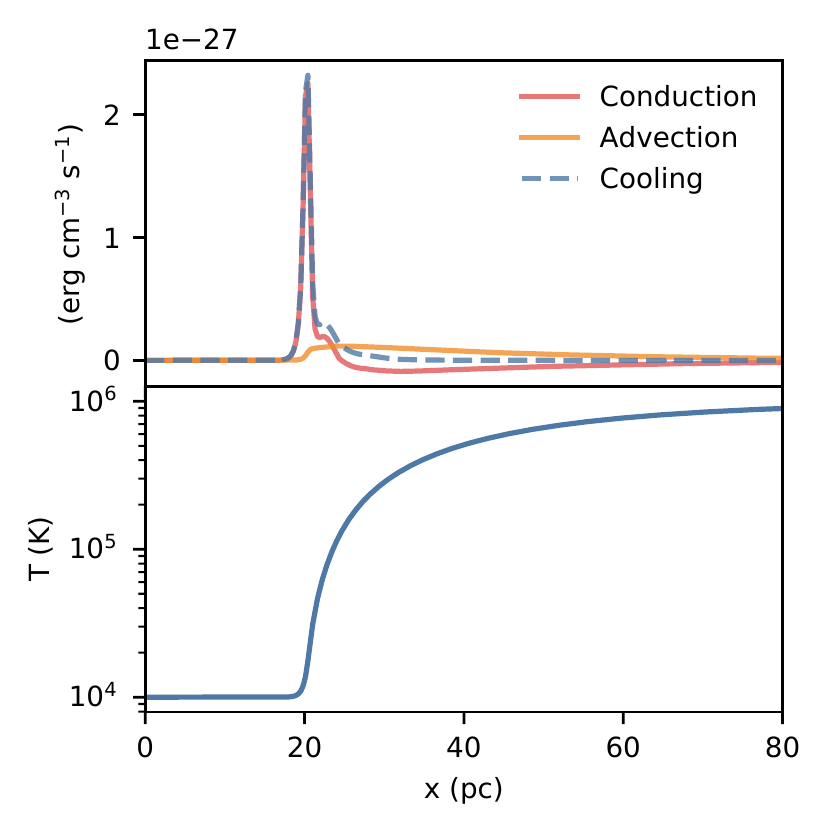}
    \caption{{\it Lower panel:} Temperature profile solution of the thermal front. {\it Upper panel:} Corresponding conductive, advective and cooling terms in Eq.~\eqref{eqn:ODE} across the thermal front.}
    \label{fig:profile}
\end{figure}

For time-steady thermal fronts, we can calculate the equilibrium solution by solving a set of coupled ODEs; this can then be compared to the time-dependent Athena\verb!++! simulations with varying resolution to understand convergence. For given boundary conditions, we can solve for the unique mass flux when we include both radiative cooling and conduction in the hydrodynamics equations \citep{kim}. We assume that $\rho v^2 \ll P$, giving us the stationary state equations in the frame of the front:
\begin{align}
    j_x &\equiv \rho v_x = \text{constant}\\
    M_x &\equiv P + \rho v_x^2 = \text{constant}\\
    \kappa \frac{\dd^2 T}{\dd x^2} &= j_x c_p \frac{\dd T}{\dd x} + \rho \mathcal{L}(T) ,
    \label{eqn:ODE}
\end{align}
where we have defined the mass flux $j_x$ and $c_p = \gamma (\gamma-1)^{-1}  k_{\rm B}/\bar{m}$ is the specific heat at constant pressure. We now have a second order ordinary differential equation in Eq.~\eqref{eqn:ODE} that can be solved numerically for the eigenvalue $j_x$, subject to the boundary conditions:
\begin{align}
    T_{-\infty} = T_1 \qquad T_{+\infty}= T_2 \qquad \frac{\dd T}{\dd x}_{\pm \infty} = 0 .
    \label{eq:BCs} 
\end{align}
Once we solve for $j_x$, we can confirm that the approximation $\rho v^2 \ll P$ holds. 

Integrating Eq.~\eqref{eqn:ODE} also yields a relationship between the mass flux and the cooling flux $Q$:
\begin{align}
    j_x = \frac{Q}{c_p (T_2-T_1)}; \ \ Q = -\int^{\infty}_{-\infty} \rho \mathcal{L} \dd x. 
    \label{eqn:mflx}
\end{align}
Equation~\eqref{eqn:mflx} makes clear that the mass flux $j_x$ depends on the detailed temperature and  density profile within  the front; thus requiring that the structure of the front be resolved. Whether the front condenses or evaporates is given by the sign of $j_x$, which in turn depends on the pressure of the system \citep{zd}. This is equivalent to a criterion on the cooling time, $t_{\rm cool} \propto P^{-1}$. At some critical pressure $P_{\rm crit}$, $j_x=0$ and the front is static. For $P > P_{\rm crit}$ ($P < P_{\rm crit}$), cooling (heating) dominates and hence there is a net mass flux from the hot (cold) phase to the cold (hot) phase. We are interested in cold gas mass growth, and so focus on the condensing case. 

From Eq.~\eqref{eqn:ODE}, we can write down two relevant length scales set by conduction \citep{kim} - the diffusion length, over which conduction balances mass flux, and the Field length, over which conduction balances radiative cooling \citep{mnb}. Figure~\ref{fig:profile} shows that the advective term  is much smaller  than the other two terms in Eq.~\eqref{eqn:ODE}, which balance one another. Thus the Field length
\begin{align}
    \lambda_{\rm F} = \sqrt{\frac{\kappa T}{n^2 \Lambda}}
\label{eq:field_length} 
\end{align}
is the relevant scale here. It was previously found in studies of thermal instability with radiative cooling that this length scale needed to be adequately resolved in order for simulations to converge \citep{koyama04,kim}. 

To verify the numerical solution for the steady front equilibrium and test for convergence with resolution, we set up the solution profile in a series of Athena\verb!++! simulations. We first initialize the simulation domain as a one dimensional box with $x=[-100,300]$\,pc. When we reduce resolution, we switch to a larger box with a range of $x=[-400,1200]$\,pc to avoid boundary effects. The front profile is generated by numerically solving the ODE for the steady state solution, and centered such that it has a temperature of $10^5$\,K at $x=0$. The left side has an initial temperature of $10^4$\,K and a number density of $10^{-2}$\,cm$^{-3}$, while the right side has an initial temperature of $10^6$\,K and a number density of $10^{-4}$\,cm$^{-3}$. These correspond to the cold and hot stable equilibrium states respectively where the net cooling rate is zero. Outflowing boundary conditions are used at both ends. With the above setup, we perform a resolution study over four orders of magnitude in order to identify what scale lengths need to be resolved in the simulation. We perform three resolution sweeps, one with the fiducial cooling curve where $\Lambda_0 = \kappa_0 = 1$, one with very strong cooling where $\Lambda_0 = 128$, corresponding to the strongest cooling used in \S\ref{sect:turb}, and one with weak conduction where $\kappa_0 = 0.1$.
By varying thermal conduction or radiative cooling at similar Field lengths, we can probe how convergence changes when the Field length is under-resolved but the relative influence of numerical and explicit diffusion is different. 

\subsection{Results} 

\begin{figure}
  \centering
  \begin{subfigure}
    \centering
    \includegraphics[width=\columnwidth]{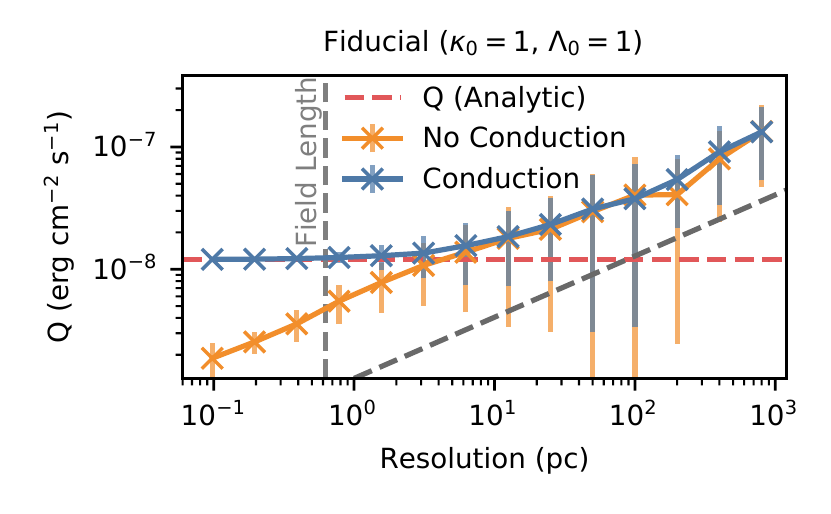}
  \end{subfigure}%
  \begin{subfigure}
    \centering
    \includegraphics[width=\columnwidth]{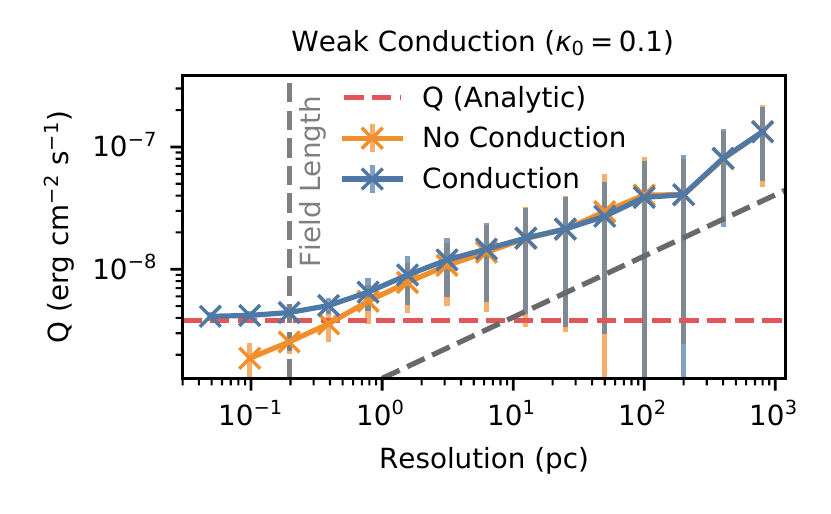}
  \end{subfigure}%
  \begin{subfigure}
    \centering
    \includegraphics[width=\columnwidth]{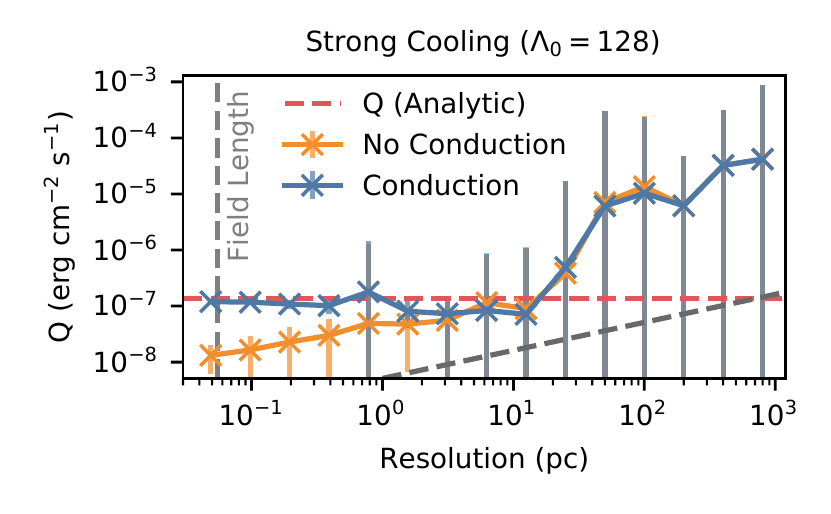}
  \end{subfigure}%
  \caption{Resolution Study: Top panel shows 1D runs with fiducial cooling and conduction. The next two panels reduce the Field length $\lambda_{\rm F}$, either by reducing conduction (middle panel), or increasing  cooling (bottom panel). While the reduced conduction case shows resolution dependence once the grid scale $\Delta > \lambda_{\rm F}$, the increased cooling case shows that $Q$ remains converged over two orders of magnitude even when $\lambda_{\rm F}$ is unresolved. Clearly, the Field length is not the key criterion when determining convergence. In the text, we argue instead that it is the relative strengths of thermal and numerical diffusion that matters. Field lengths, analytic solutions and $\sqrt{\Delta}$ scalings (grey dashed lines) are also shown for comparison.}
  \label{fig:resolution}
\end{figure}

The bottom panel of Fig.~\ref{fig:profile} shows the temperature profile of the solution obtained from the ODE in Eq.~\eqref{eqn:ODE}, which we solve via the shooting method. The final solution corresponds to an inflow velocity of 2 ${\rm km \, s}^{-1}$ in the hot gas. The top panel of Fig.~\ref{fig:profile} shows the relative importance of the conductive and advective terms in Eq.~\eqref{eqn:ODE}; conduction balances cooling over most of the front. By varying the parameters $\kappa_0$ and $\Lambda_0$, we find that $j_x,Q \propto (\kappa/t_{\rm cool})^{1/2}$, as expected from Eqs.~\eqref{eqn:mflx} and \eqref{eq:field_length}. This provides a reference solution which we compare against the Athena\verb!++! results in our resolution study. 

The results of the resolution study are shown in the top panel of Fig.~\ref{fig:resolution}. First consider the runs with conduction. With increasing resolution, we see convergence towards the mass flux computed from Eq.~\eqref{eqn:mflx}. At high resolution, when the Field length is resolved, the structure of the thermal front is resolved and agrees with the reference solution. However, as we lower the resolution, $Q$ deviates from the reference solution, and increases steadily $Q \propto \sqrt{\Delta}$ (where $\Delta$ is the grid scale).These are marked by dashed lines in Fig.~\ref{fig:resolution}. In the runs with no thermal conduction, only numerical diffusion balances cooling. Convergence vanishes and throughout the entire range, $Q \propto \sqrt{\Delta}$; as $\Delta \rightarrow 0$, $Q \rightarrow 0$. This is in line with the expectation that for zero conduction, there should be a vanishing mass flux. All of the behavior in the fiducial case is in agreement with canonical expectations. 

Since -- in line with previous expectations -- the Field length $\lambda_{\rm F} \propto \sqrt{\kappa t_{\rm cool}}$ appears to be the critical scale which must be resolved, we reduce it in two ways, either by reducing the cooling time $t_{\rm cool}$ or reducing the conductivity $\kappa$. We find that these two procedures do not give the same result for the same reduced Field length. If we keep cooling fixed but reduce conduction (middle panel of Fig.~\ref{fig:resolution}), then $Q$ becomes resolution dependent once $\Delta \gtrsim \lambda_{\rm F}$, as expected. 
By contrast, in the setup with strong cooling, (bottom panel of Fig.~\ref{fig:resolution}), we find that the mean cooling $Q$ is slightly lower but still close to the converged value for $\Delta \lesssim 100 \lambda_{\rm F}$, even though the Field length is completely unresolved. Instead, lower resolution results in rapid temporal oscillations in $Q$, which increase in amplitude for lower resolution. Instead of an offset, $Q$ simply oscillates about the correct equilibrium value. Similar behavior is also observed in the 3D simulations as described further below in \S\ref{sect:3D_convergence} and shown in Fig.~\ref{fig:3d_resolution}. Here, we plot the mean value of $Q$, while error bars indicate the standard deviation.

These results make clear that one must distinguish between errors due an unresolved front (stiff source terms) and errors due to numerical diffusion. In our case, an unresolved front contributes to the variance of the solution (numerical dispersion), but does not bias the solution. It can be beaten down by time averaging. Numerical diffusion, on the other hand, unavoidably biases the solution. The criterion for a converged solution is therefore not $\Delta < \lambda_{\rm F}$, but rather $D_{\rm num} < D_{\rm thermal}$; i.e. that numerical diffusion is subdominant to thermal diffusion.

To expand on this point: the static radiative interface is a stiff problem where the source term (radiative cooling)  defines a length scale (the Field length, over which thermal diffusion and radiative cooling  balance) which is often much smaller than other scales of interest and can lie below the grid scale. It is well-known that hyperbolic systems with a stiff source term which is  unresolved can have wave speeds which are either spurious (e.g., see \citealp{colella86} for detonation waves), or  still centered about the correct value, albeit with  a larger dispersion (e.g., \citealp{leveque2002}, see \S17.10-17.18). Relaxation systems are known to be well-behaved if certain subcharacteristic requirements are satisfied; although the reason is still not fully understood \citep{pember93}. At least with Athena\verb!++!, which uses a stable, second-order accurate  modified Gudunov method for handling stiff source terms \citep{sekora10}, and the two-moment conduction module we have used, radiative thermal fronts appear  to fall into this class of problem, potentially because the sound speed of the cold gas sets a characteristic velocity scale. When the  Field length  is not  resolved, cooling and conduction cannot balance exactly due  to discretization errors in the temperature and its derivatives. Instead, they (and  hence $Q$) oscillate  about thermal balance and the true answer. While numerical  diffusion creates systematic biases in the true steady  state solution, numerical dispersion creates fluctuating errors which can be  averaged out over a long time series. 
Of course, also the latter can only buy a limited amount of dynamic range  before errors swamp the solution (in the example shown in the lower panel of Fig.~\ref{fig:resolution}, it is $\sim 2$ orders of magnitude). We quantify this effect below.

\begin{figure}
    \centering
    \includegraphics{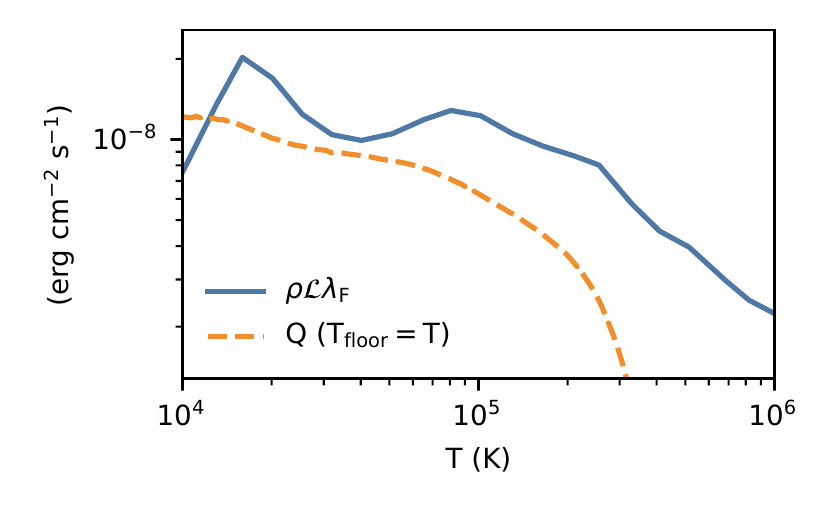}
    \caption{The distribution of net cooling with temperature. The solid line shows the integrand of Eq.~\eqref{eq:Q_T}, while the dashed line shows $Q(T_{\rm floor}=T)$, where the lower limit of the integral in Eq.~\eqref{eq:Q_T} is clipped at $T_{\rm floor}$.  Note that $Q$ has contributions from a broad range of temperatures from $10^{4} {\rm K} < T < 10^{5} \, {\rm K}$, and does not plummet until $T_{\rm floor} > 10^{5} \, {\rm K}$.}
    \label{fig:Q_integrand}
\end{figure}

The  Field length is a strong function of temperature, and despite the fact that (under isobaric conditions) volumetric emissivity  peaks at few~$\times 10^{4}$~K, the contribution to the integral surface brightness is more broadly distributed:
\begin{align}
    Q =& \int \rho \mathcal{L} \dd x   =  \int \rho \mathcal{L} \frac{T}{T^{\prime}} \frac{\dd T}{T} \nonumber\\ 
    \approx&  \int \rho \mathcal{L} \lambda_{\rm F} \dd(\log T)  \propto \left( \frac{\Lambda(T)  \kappa}{T} \right)^{1/2} ,
    \label{eq:Q_T} 
\end{align}
where we have used $nT \approx$~const and $T' \approx T/\lambda_{\rm F}$. Figure~\ref{fig:Q_integrand} shows the integrand $\rho \mathcal{L} \lambda_{\rm F}$. It has two distinct peaks at $T \sim 10^{4}$ and $10^{5}\,$K  for the $\kappa=$~const case considered here, and is dominated by higher temperatures $T  > 10^{5}$K for the more realistic case of Spitzer conduction. Even though  the volumetric emissivity peaks at $T \sim 10^{4}$K, Q has contributions from a broad range of temperatures, because the Field length is a strongly increasing function of temperature:   
\begin{equation}
    \lambda_{\rm F} = \left( \frac{\kappa T}{n^{2} \Lambda(T)} \right)^{1/2} \propto \left( \frac{T^{3+n}}{\Lambda(T)}\right)^{1/2} ,
\end{equation}
where $\kappa \propto T^{n}$ (and $n=0$ for $\kappa=$ const, $n=5/2$ for Spitzer conduction).  Thus, even when  $\lambda_{\rm F}(T\sim  10^{4}\,\mathrm{K})$  is unresolved, $Q$ will be approximately correct as long as $\lambda_{\rm F}(T\sim  10^{5}\,\mathrm{K})$ is resolved. This explains well the numerical results shown in  the lower panel of Fig.~\ref{fig:resolution}, for which $\lambda_{\rm F}(T\sim 10^5\,\mathrm{K}) / \lambda_{\rm F}(T\sim 10^4\,\mathrm{K}) \sim 30$.

In summary: in our 1D simulations, in the absence of thermal conduction, the surface brightness is resolution dependent $Q \propto \sqrt{\Delta}$. If explicit thermal  conduction  is included and larger than numerical diffusion, then $Q$ is numerically  converged, even  if the Field  length is unresolved. The unresolved Field  length merely contributes to an increased variance. However, once $\Delta > \lambda_{\rm F} (T\sim 10^{5} K) \sim 30 \lambda_{\rm F}(T \sim 10^{4} K)$, the error bars grow  rapidly. 

\dobib

%% file: sections/4_combustion.tex
\section{Analytic Estimates from Turbulent Combustion} \label{sect:combustion} 

In 3D, turbulence in the mixing layer complicates matters considerably. In this section, we explore parallels between radiative and combustion fronts, and review findings from the turbulent combustion literature. Based on this, we also develop an analytic model of radiative TMLs.

\begin{figure*}
    \centering
    \includegraphics[width=.75\textwidth]{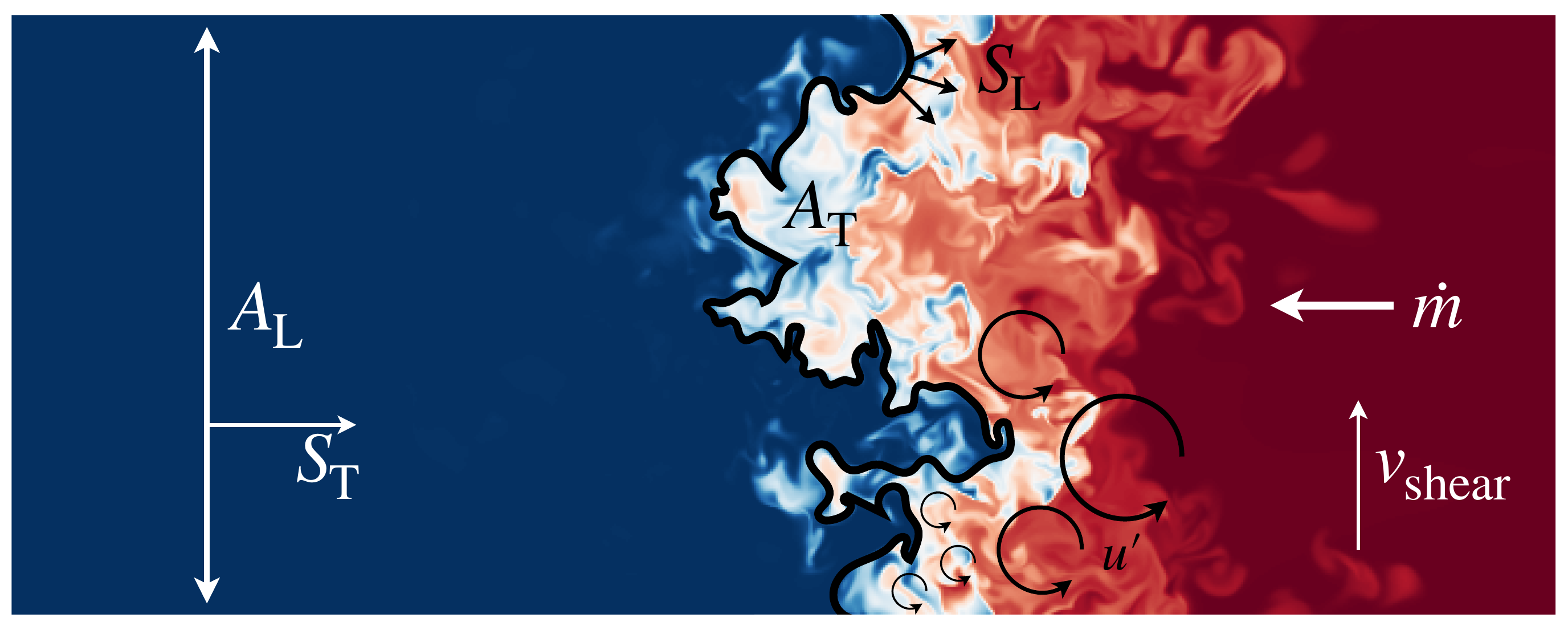}
    \caption{Slice through a mixing layer simulation with the relevant quantities          from combustion theory marked.}
    \label{fig:mixinglayer_sketch}
\end{figure*}

\subsection{Introduction \& Terminology}
There are close parallels between a two-phase radiative front and a combustion front. In a radiative front, the `fuel' is hot gas, which `burns' (i.e. cools radiatively) in a temperature and density dependent manner upon mixing with `oxidizer' (cold gas) to produce `ash' (more cold gas). The one unusual characteristic in radiative fronts is that the end product is more `oxidizer'. Moreover, combustion fronts share the property that the rate of burning and hence speed of front propagation is determined by conditions within the front, which in general must therefore be resolved; it similarly reduces to a non-linear eigenvalue problem in 1D \citep{zeldovich85}. Due to the obvious bearing of turbulent combustion on gas chamber combustion engines in automobiles and jet engines, with implications for fuel efficiency and air pollution, the literature is vast (see \citealp{kuo12} for a recent comprehensive textbook). Importantly, besides sophisticated high resolution numerical simulations, there is a plethora of experimental results. Within the astrophysical community, similar issues arise in thermonuclear burning fronts on carbon-oxygen white dwarfs, where conditions within the front determine the burning rate. The large scale separation ($\sim10^{7}$) between the size of the white dwarf (which sets the outer scale of turbulence) and the front structure precludes direct calculation of the fronts in simulations; a subgrid model (e.g., \citealp{niemeyer95, schmidt06, jackson14}) is necessary. Here, we draw upon this literature to provide an intuitive physical picture for the puzzles described in \S\ref{sect:intro}. 

In the language of combustion literature, in radiative TMLs the fuel and oxidizer are not perfectly pre-mixed before combustion. Instead, the two are initially separate. They are subsequently entrained and dispersed by large scale eddies with turbulent velocity $u^\prime$. The turbulent cascade down to small scales leads to stretching, fragmentation and a vast increase in surface area, until at small scales diffusion mixes the fuel and oxidizer and allows combustion to take place. The net result is that fuel is consumed at a rate $\dot{m}$. The process is similar to how stirring enables a vast increase in surface area and the large mixing rate between coffee and cream, despite the long molecular diffusion time.

What is the characteristic velocity at which a thermal front propagates, and how do results depend on the source of thermal diffusion? Thermal diffusion is canonically due to thermal conduction. In the absence of turbulence, this gives rise to a `laminar flame speed' $S_{\rm L} \sim \sqrt{D_{\rm L}/t_{\rm cool}}$, where $D_{\rm L}$ is the customary diffusion coefficient with  units of $L^{2} T^{-1}$. This can be seen by balancing thermal diffusion with cooling (Eq.~\eqref{eqn:mflx} below), or simply from dimensional analysis. If, however, thermal conduction is not included, which is often the case in many numerical simulations, then all diffusion is numerical: gas mixing and thermal diffusion operate close to the grid scale. When numerical diffusion dominates, $D \sim v \Delta x$, where $v$ is a characteristic velocity and $\Delta x$ is the grid scale. Thus, in the absence of conduction, $S_{\rm L} \propto \sqrt{\Delta x}$. 

Turbulence gives rise to a large increase in surface area of the phase boundary $A_{\mathrm{T}}\gg A_{\mathrm{L}}$ which leads to a `turbulent flame speed' $S_{\rm T}\gg S_{\rm L}$, where $S_{\rm T}$ (corresponding to $v_{\rm in}$ in Eq.~\eqref{eq:vin}) characterizes the rate at which fuel is consumed and the overall front propagates. Our goal is to understand $S_{\rm T}$, which sets the hot gas entrainment rate. In Fig.~\ref{fig:mixinglayer_sketch} we visualize the quantities introduced in this section. 

In this paper, we consider the impact of shear driven turbulence on the radiative front. Evaporative fronts are subject to a well-known corrugational instability, the Darrieus-Landau instability \citep{landau87,zeldovich85,inoue06}, which deforms the front and drives turbulence, which can also lead to increased surface area and accelerated reaction rates. Here, we focus only on condensation fronts. 

\subsection{Characteristic Regimes}
Turbulent combustion is characterized by several dimensionless numbers \citep{kuo12}. These give rise to classification into many distinct burning regimes which are typically shown on a plot known as a Borghi diagram. For our purposes, the most useful parameter is the \Da number: 
\begin{equation}
    {\rm Da} = \frac{\tau_{\rm turb}}{\tau_{\rm react}} = \frac{L}{u^{\prime} t_{\rm cool} (T)},
    \label{eq:Da}
\end{equation}
which gives the ratio of the eddy turnover time at the outer scale $L$, with turbulent velocity $u^\prime$, to a reaction time\footnote[2]{Another important parameter in combustion studies is the Karlovitz number ${\rm Ka} = \delta^{2}/\eta^{2}$, where $\delta$ is the diffusive scale (corresponding to the Field length) and $\eta$ is the Kolmogorov scale at which viscosity damps turbulence. It determines whether the propagation of the small scale interface is set by laminar burning, or whether turbulence alters the structure of the diffusive front. Since we do not have explicit thermal conduction or viscosity, the flamelet fronts are generally unresolved and we have $\delta \sim \eta \sim \Delta x$, i.e. ${\rm Ka} \sim 1$ in our simulations.}. For us, it can also be viewed as ${\rm Da} \sim L/L_{\rm cool}(T)$, the ratio of the integral length scale of turbulence to the cooling length $L_{\rm cool} (T) \sim u^{\prime} t_{\rm cool}(T)$.
Note that the cooling time $t_{\rm cool}(T)$ varies with temperature through the front. 
The \Da number separates two asymptotic regimes, $\mathrm{Da} \ll 1$ (`well-stirred reactor') and $\mathrm{Da} \gg 1$ (`corrugated flamelets') which are universal across all classification schemes. 
The cooling time $t_{\rm cool}(T)$ decreases continually across the front, as the temperature $T$ declines from the hot to the cold gas temperature ($T_{\rm h}$ and $T_{\rm c}$, respectively) and the cooling function peaks extremely close to $T_{\rm c}$. Initially, close to the hot gas boundary, $\mathrm{Da} < 1$. Turbulence cascades to the diffusion (grid) scale before the two components react. Fuel and oxidizer are well mixed and thus the reaction rate is uniform across the entire volume. This is known as the `well-stirred reactor' regime. 
In mixing length theory, this can be characterized by a turbulent diffusion coefficient $D_{\rm turb} \sim u^{\prime} L$. 
Thus, in the weak cooling (${\rm Da} < 1$) regime, we expect: 
\begin{equation}
  S_{\rm T} \approx \left(\frac{D_{\rm turb}}{t_{\rm cool}} \right)^{1/2} \approx \left(\frac{u^{\prime}L }{t_{\rm cool}} \right)^{1/2}. 
  \label{eq:vin_weak_cooling}
\end{equation}

However, as we move toward the cold gas boundary, the temperature and cooling time fall. When ${\rm Da} > 1$, burning proceeds before mixing is complete, and combustion thus takes place inhomogeneously. In our context, inhomogeneous cooling leads to fragmentation into a multiphase medium. The criterion ${\rm Da} \sim 1$ corresponds to the transition between single and multiphase structure in the mixing layer. The steep temperature dependence of cooling means that most cooling takes place in thin unresolved fronts close to $T \sim 10^{4}$K at the interface between cold and hot gas. The turbulent cascade wrinkles this interface and vastly increases its area, thus increasing the volumetric cooling rate. In a famous paper, \citet{damkohler40} conjectured that the increase in surface area leads to a turbulent flame speed: 
\begin{equation} 
    S_{\rm T} \approx S_{\rm L} \left( \frac{A_{\rm T}}{A_{\rm L}} \right),
    \label{eq:area_ratio} 
\end{equation}
where $A_{\rm T}$ and $A_{\rm L}$ are the turbulent and laminar flame areas. This comes from simply equating the mass flux through $A_{\rm L}$ at velocity $S_{\rm T}$ with the mass flux through $A_{\rm T}$ at velocity $S_{\rm L}$, as illustrated in Fig.~\ref{fig:mixinglayer_sketch}. This intuitive notion can be made more precise and proven \citep{bray91}.

\subsection{Scalings of the turbulent velocity $S_{\rm T}$}
To calculate $S_{\rm T}$, we therefore need to know $A_{\rm T}/A_{\rm L}$. This has no single consensus answer; for instance, Table 5.1 of \citet{kuo12} lists 20 fits to $S_{\rm T}/S_{\rm L}$ obtained from theory, simulation and experiment depending on geometry, boundary conditions and flame wrinkling process. One simple way of parametrizing most known scalings is to write: 
\begin{equation}
    \frac{S_{\rm T}}{S_{\rm L}} = \frac{A_{\rm T}}{A_{\rm L}} = 1 + \left( \frac{u^{\prime}}{S_{\rm L}}  \right)^{n} \approx \left( \frac{u^{\prime}}{S_{\rm L}}  \right)^{n},
    \label{eqn:n-scaling} 
\end{equation}
where the last equality holds for $u^{\prime} \gg S_{\rm L}$.
The most well-known scaling is $n=1$ \citep{damkohler40}, which has substantial experimental support in a variety of settings. For instance, \citet{libby79,clavin79,peters88,bray90,bedat95} obtain similar scalings in both theory and experiment. It implies 
\begin{equation}
    S_{\rm T} \approx u^{\prime} ,
    \label{eq:damkohler} 
\end{equation} 
i.e. that the combustion front simply propagates at the turbulent velocity. A useful geometrical interpretation comes from \citet{damkohler40} and \citet{shchelkin43} who considered the distortion of the flame-burning front into several `Bunsen cones' -- analogous to a Meker burner.
A simplified version of his argument is as follows: consider a flat interface of area  $A_{\rm L} = L^{2}$. It propagates in a direction normal to  the front at velocity $S_{\rm L}$. Over  a burning  time $t_{\rm burn} \sim L/S_{\rm L}$, laminar burning will traverse a distance $L$, whereas turbulent motions traverse a distance $l_{\rm turb} \sim u^{\prime}   t_{\rm burn} \sim (u^{\prime} /S_{\rm L})  L$, creating a wrinkled (conical) region with area $A_{\rm T} \sim l_{\rm turb} L \sim (u^{\prime} /S_{\rm L}) A_{\rm L}$. Thus, $A_{\rm T}/A_{\rm L} \sim u^{\prime}/S_{\rm L}$. A more careful consideration of the conical geometry gives
\begin{equation} 
    \frac{A_{\rm T}}{A_{\rm L}} \sim \left( 1+ \left(\frac{2 u^{\prime}}{S_{\rm L}} \right)^{2} \right)^{1/2},
\end{equation}
which  reduces to $A_{\rm T}  \propto  u^{\prime}$  for $u^{\prime} \gg S_{\rm L}$. Note that in our context, $S_{\rm T} \sim u^{\prime}$, independent of all other parameters, including $S_{\rm L}$, which in general is resolution dependent. 

Thus far, we have ignored the influence of other parameters. 
As previously mentioned, turbulent combustion is in fact characterized by at least two dimensionless numbers in a Borghi diagram, typically either $({\rm Re}_{\rm L}, {\rm Da})$ or $(\eta/\delta, u^{\prime}/S_{\rm L})$, where $\delta$ is the thermal diffusive scale and $\eta$ is the Kolmogorov scale. An important boundary in the Borghi diagram is the Klimov-Williams line, where ${\rm Ka} = (\delta/\eta)^{2} \sim  1$, where  laminar  flame scales and turbulent stretching scales become comparable. As noted earlier, in numerical codes where numerical diffusion is dominant, we expect $\delta \sim \eta \sim \Delta$, so that ${\rm Ka} \sim 1$, and we are always in this regime. In a broad neighborhood of the Klimov-Williams line, flame propagation has been argued to obey  the scaling \citep{gulder91}: 
\begin{equation}
    \frac{S_{\rm T}}{u^{\prime}} = {\rm Da}^{1/4} = \left(\frac{L}{u^{\prime}{\tau_{\rm react}}} \right)^{1/4},
    \label{eq:gulder91} 
\end{equation} 
which fits a large body of burning velocity data  \citep{gulder91,zimont95}. Note that this is precisely the $v_{\rm mix} \approx c_{\rm s,cold} (t_{\rm cool}/t_{\rm sc,cold})^{-1/4}$ scaling previously reported \citep{gronke18,gronke19,fielding20}
if we identify  $u^{\prime} \approx c_{\rm s,cold}$ and ${\rm Da} \sim L/(u^{\prime} t_{\rm react})  \sim t_{\rm sc,cold}/t_{\rm cool}$ -- but backed up by experimental data.
In this work, we will test and confirm the resolution independence of $v_{\rm in}$ (cf. \S\ref{sect:3D_convergence}), and thus continue with Eq.~\eqref{eq:gulder91} as our `fiducial' scaling in the strong cooling regime where Da > 1.
Note that while combustion theory can provide a link between $S_{\mathrm{T}}$ and $u'$ and helps us understand the core questions presented in \S\ref{sect:intro}, the scaling of $u'$ with respect to the flow properties depend on the turbulent driving process and have to be found from numerical experiments (see \S\ref{sect:uprime_scalings}).

\subsection{Details of the Fiducial $S_{\rm T}/u^{\prime} = {\rm Da}^{1/4}$ Scaling}
\label{sect:fiducial_scaling} 

What is the theoretical justification for Eq.~\eqref{eq:gulder91}? Following \citet{tennekes68,kuo72}, \citet{gulder91} argues that turbulent vortex tubes should be separated by a distance of order the Taylor microscale\footnote[3]{The Taylor microscale is a lengthscale which comes from a Taylor series expansion of flow correlations; it is the scale at which shear is maximized. While it is larger than the Kolomogorov scale, it can be thought to demarcate the end of the inertial range and the beginning of the dissipation range.}. Assuming that laminar burning fronts must cover a distance of order the Taylor microscale to complete burning, he arrives at Eq.~\eqref{eq:gulder91}. We do not recount his arguments here, but instead refer interested readers to the original paper.

It is not clear how applicable the \citet{gulder91} argument is to our numerical simulations, which do not have explicit viscosity and a well defined Reynolds number, and thus do not have a well-defined Taylor microscale (which will vary with resolution). For us, the most important fact is that there is significant experimental evidence in turbulent combustion data for the scaling in Eq.~\eqref{eq:gulder91}, which do not suffer from the same limitations as our numerical simulations. Here, we propose a simpler alternative argument which gives similar results. 

What is the effective cooling time $\tilde{\tau}_{\rm cool}$ of an inhomogeneous medium where ${\rm Da} > 1$? It is clearly not the standard cooling time $t_{\rm cool}$, since only a small fraction of the medium is cooling. Consider cold gas as a scalar pollutant, which diffuses over scales $\lambda$ on a timescale $t_{\rm D} \sim \lambda^{2}/D_{\rm turb}$. In the fast cooling (Da > 1) limit, all of the mixed gas will cool. Over a cooling time, the cold gas diffuses over a distance (setting $t_{\rm D} \sim t_{\rm cool}$): 
\begin{equation} 
\lambda_{\rm cool} \sim \sqrt{D_{\rm turb} t_{\rm cool}} \sim \sqrt{L u^{\prime} t_{\rm cool}} 
\label{eq:turb_field} 
\end{equation}  
Thus, after a cooling time $t_{\rm cool}$, only a fraction $f_{\rm cool} \sim \lambda_{\rm cool}/L$ of the gas in an eddy has mixed and cooled; only after $N_{\rm cool} \sim f_{\rm cool}^{-1} \sim L/\lambda_{\rm cool}$ cooling times does all the gas in the eddy cool. The effective cooling time is therefore: 
\begin{equation}
    \tilde{\tau}_{\rm cool} \sim N_{\rm cool} t_{\rm cool} \sim \frac{L}{\lambda_{\rm cool}} t_{\rm cool} \sim \sqrt{\frac{L}{u^{\prime}} t_{\rm cool}},  
\label{eq:tilde_tau} 
\end{equation}
i.e., the geometric mean of the eddy turnover time and the cooling time. Equivalently, we can view Eq.~\eqref{eq:turb_field} as the effective mean free path of a fluid element. The mean free time is therefore:
\begin{equation} 
    \tilde{\tau}_{\rm cool} \sim \frac{\lambda_{\rm cool}}{u^\prime} 
    \sim \sqrt{\frac{L}{u^{\prime}} t_{\rm cool}}, 
\end{equation}
which gives the same result. Hot gas in the multiphase, strong cooling region is converted to cold gas on a timescale $\tilde{\tau}_{\rm cool}$, which is shorter than the mixing time $L/u^{\prime}$, but longer than than homogeneous cooling time $t_{\rm cool}$, since only a small fraction of the volume is cooling. 

Equation~\eqref{eq:tilde_tau} is a common random walk result.  For instance, consider a photon in a medium which scatters (with optical depth $\tau_{s}$) and absorbs (with optical depth $\tau_{a}$). Then the effective optical depth is $\tau_{*} \sim \sqrt{\tau_a \tau_s}$, with effective survival time $t_* \sim \sqrt{t_{a} t_{s}}$ \citep{rybicki79}. Similarly, when considering the competition between thermal conduction and cooling, the Field length: 
\begin{equation}
    \lambda_{\rm F} \sim \sqrt{\frac{\kappa T}{n^{2} \Lambda(T)}} \sim \sqrt{\lambda_{e} v_{e} t_{\rm cool}}   
\end{equation}
(using $\kappa \sim P v_{e} \lambda_{e}/T$) is the geometric mean of the elastic ($\lambda_{\rm e}$; the Coulomb mean free path) and inelastic ($v_{e} t_{\rm cool}$) mean free paths for a thermal electron. Considering the Field length as the effective mean free path for an electron, the mean free time is $t_{e} \sim \lambda_{F}/v_{e} \sim \sqrt{t_e t_{\rm cool}}$, where $t_{e} \sim \lambda_{\rm e}/v_e$. Eqs.~\eqref{eq:turb_field} and \eqref{eq:tilde_tau} are the equivalent analogs for a turbulent eddy, with $\lambda_{e} \rightarrow L$, $v_e \rightarrow u^{\prime}$. Note that the largest eddies dominate mixing, and so therefore all quantities related to turbulence are evaluated at the outer scale $L$. Later in \S\ref{sect:mixing_length}, we shall see from simulation results that (as assumed here) the turbulent diffusion coefficient $D_{\rm turb} \sim u^{\prime} L$ is relatively unaffected by cooling.   

Substituting $\tilde{\tau}_{\rm cool}$ (Eq.~\eqref{eq:tilde_tau}) for the cooling time in the usual expression for the turbulent flame velocity (Eq.~\eqref{eq:vin_weak_cooling}), we obtain: 
\begin{equation}
    S_{\rm T} \sim \left( \frac{D_{\rm turb}}{\tilde{\tau}_{\rm cool}} \right)^{1/2} \sim u^{\prime} \left( \frac{L}{u^{\prime} t_{\rm cool}} \right)^{1/4} \sim u^{\prime} {\rm Da}^{1/4}.
    \label{eq:fiducial-scaling}
\end{equation}

Beyond the turbulent flame speed (Eq.~\eqref{eq:fiducial-scaling}), this ansatz makes predictions which are testable in the simulations: 
\begin{itemize} 

\item{{\it Effective emissivity.} This model predicts an effective cooling time in the multiphase region given by Eq.~\eqref{eq:tilde_tau}, so that the effective emissivity is: 
\begin{equation}
    \tilde{\epsilon} \sim \frac{P}{\tilde{\tau}_{\rm cool}} \sim P \left( \frac{u^{\prime}}{t_{\rm cool} L} \right)^{1/2}.
    \label{eq:emissivity_eff} 
\end{equation}
The $\tilde{\epsilon} \propto {u^{\prime}}^{1/2}t_{\rm cool}^{-1/2}$ scaling can be checked in the simulations.}

\item{{\it Width of multiphase regions.} Equivalently, if Eqs.~\eqref{eq:fiducial-scaling} and \eqref{eq:emissivity_eff} hold, then we can use $Q \sim P v_{\rm in} \sim \tilde{\epsilon} h$ to find that the width $h$ of the multiphase region scales as: 
\begin{equation}
    h \propto L \left( \frac{u^{\prime} t_{\rm cool}}{L} \right)^{1/4} \propto {\rm Da}^{-1/4} ,
    \label{eq:h_fast_cooling} 
\end{equation}
where the $h \propto (u^{\prime})^{1/4} t_{\rm cool}^{1/4}$ scaling can be tested in the simulations. Of course, of Eqs.~\eqref{eq:fiducial-scaling}, \eqref{eq:emissivity_eff}, and \eqref{eq:h_fast_cooling}, only two are independent. 
}

\end{itemize}

We caution once again that there does not appear to be universally applicable turbulent velocity scalings in the literature, which tend to be situation dependent.  Nonetheless, it is reassuring to see that the scalings we see in our numerical simulations with limited dynamic range have also been seen in a large body of experimental data and have theoretical justification. 

\subsection{Implications for the fractal nature of mixing layers}
These properties can also be related to the fractal nature of radiative mixing layers. Recently, \citet{fielding20} showed that the area of the cooling surface in radiative mixing layer simulations obeys a fractal scaling, with 
\begin{equation}
    \frac{A_{\rm T}}{A_{\rm L}} = \left( \frac{\lambda}{L} \right)^{2-D},
    \label{eq:area_scaling} 
\end{equation}
where $\lambda$ is the smoothing scale and $D=2.5$ was the fractal dimension argued to hold by analogy with well-known fractals, and verified in their simulations. Turbulence combustion fronts are indeed well known to be fractals, due to the dynamical self-similarity of turbulence in the inertial range. Experimental measurements by e.g. instantaneous laser tomography have given values ranging from $D=2.1-2.4$ in a variety of flow geometries, with a preferred value of $D=2.35$ \citep{hentschel84,sreenivasan89}; it has been argued that this fractal dimension is universal \citep{catrakis02,aguirre05}. From Eq.~\eqref{eq:area_ratio}, the fractal dimension can be used to calculate the turbulent flame speed \citep{gouldin86,peters88}. The fractal scaling and consequent increase in area $A_{\rm T}$ should extend all the way down to the Gibson scale $\lambda_{\rm G}$, which is defined to be the scale where the turbulent velocity equals the laminar flame speed, $v(\lambda_{\rm G}) = S_{\rm L}$. This is often unresolved in simulations. If we use the Kolmogorov scaling $v \propto \lambda^{1/3}$, then we obtain: 
\begin{equation}
    \frac{S_{\rm T}}{S_{\rm L}} = \frac{A_{\rm T}}{A_{\rm L}} = \left( \frac{\lambda_{\rm G}}{L} \right)^{2-D} = \left( \frac{u^{\prime} }{S_{\rm L}} \right)^{3(D-2)} ,
\end{equation}       
where we have used Eq.~\eqref{eq:area_scaling} and $v(\lambda_{\rm G}) = S_{\rm L}$. Thus, in Eq.~\eqref{eqn:n-scaling}, we have $n=3(D-2)$. The experimental value of $D=2.35$ gives $n=1.05$, in good agreement with Damk{\"o}hler's scaling, and fair agreement with the scaling in Eq.~\eqref{eq:gulder91}. The \citet{fielding20} value of $D=2.5$ gives $n=1.5$, or $S_T = u'(u'/S_L)^{1/2}$. If one uses the laminar $S_L \propto t_{\rm cool}^{-1/2}$ from our static simulations, this would imply $S_T \propto t_{\rm cool}^{1/4}$. However, in the \citet{fielding20} model, the speed at which a cooling layer advances is $S_L \propto t_{\rm cool}^{1/2}$, so they end up with $S_T \propto t_{\rm cool}^{-1/4}$ as well. The scalings are sensitive to the fractal dimension $D$ and the measurement error on $D$ obtained from the simulations is unclear at this point. In addition, the cutoff scale of turbulence may not be the Gibson scale. We caution that fractal arguments have not proven to be fully robust in the turbulent combustion context. For instance, the measured fractal parameters fluctuate depending on the extraction algorithm, and have not been able to correctly predict the turbulent burning velocity \citep{cintosun07}. 

\subsection{Implications for energetics and convergence criteria}

The above considerations bear upon the two over-arching questions first raised in \S\ref{sect:intro}, which will be further addressed in the course of this paper. 

{\it Energetics.} Why is $S_{\rm T} \sim c_{\rm s,cold}$? From Eq.~\eqref{eq:damkohler}, we have $S_{\rm T} \sim u^{\prime}$, i.e. of order the turbulent velocity at the outer scale.
The timescale of the Kelvin-Helmholtz instability, which mixes the two fluids, is $t_{\rm KH} \sim \sqrt{\chi} L/v_{\rm shear}$; the characteristic turbulent velocity of the interface between hot and cold gas is $u^{\prime} \sim v_{\rm shear}/\sqrt{\chi} \sim \mathcal{M}_{\rm hot}  c_{\rm s,hot} /\sqrt{\chi} \sim \mathcal{M}_{\rm hot} c_{\rm s,cold}$. If $\mathcal{M}_{\rm hot} \sim 1$, as is true for many situations in the CGM (since the virial velocity is of order the virial sound speed), this reduces to $u^{\prime} \sim c_{\rm s,cold}$. We will study detailed scalings of $u^{\prime}$ in \S\ref{sect:uprime_scalings}. 

{\it Resolution independence.} Neither our fiducial scaling (Eq.~\eqref{eq:fiducial-scaling}) nor the \Da scaling (Eq.~\eqref{eq:damkohler}) depend on the diffusion coefficient, and thus are independent of resolution. Physically, this is because most radiative cooling takes place in the ${\rm Da} > 1$ regime, when the cooling time is shorter than the eddy turnover time. When cooling is `fast' compared to mixing, all gas which mixes cools -- the rate limiting step is the rate at which turbulence cascades to diffusive scales, whereupon mixing and cooling happen on very short timescales. The time the turbulent cascade takes to reach small scales is simply $\tau_{\rm turb} \sim L/u^{\prime} $ the eddy turnover time at the outer scale, since in Kolmogorov turbulence, the eddy turnover time $\tau_{\rm l} \sim l/v_{\rm l} \propto l^{2/3}$ is a progressively smaller function of scale. The situation is similar to passive scalar mixing, except that here the passive scalar which is being advected is temperature. The rate at which coffee mixes with cream is given by the stirring time of the spoon, independent of the details of molecular diffusion. Similarly, the rate at which hot gas mixes with cold gas and subsequently cools is given by the eddy turnover time at the outer scale, independent of the details of thermal (numerical) diffusion, which set the structure of the (often unresolved) laminar thermal fronts. Thus, the important scale that needs to be resolved is the mixing due to turbulent eddies at the outer scale.

\dobib

%% file: sections/5_3D.tex
\section{3D Simulations: Turbulent Fronts} \label{sect:turb}

We next turn to 3D simulations of radiative mixing layers. Due to an additional ingredient -- turbulence -- not present in 1D  simulations, their properties are quite different. In this section, we compare the results of 3D simulations to the model discussed in the previous section.

\subsection{Setup} \label{sect:3Dsetup} 

Our setup closely follows the work of \citet{ji19}. The coordinate system is as follows:  $y$ is the axis of shear flow, $x$ is normal to the cold/hot interface  (the principal direction of interest along which front  properties vary), and $z$ is the third remaining dimension. Boundary conditions are periodic along the $y$ and $z$ axes and outflowing along the $x$ axis. The bounds of the $x$ axis are $[-100,\,200]$~pc and the bounds for the $y$ and $z$ axes are $[0,100]$~pc. Cold $10^4$~K gas is initially located in the negative $x$ region and hot $10^6$~K gas in the positive $x$ region, separated by a smoothly varying front centered at $x=0$ where $T = 10^5$~K. The initial front profile is obtained by solving for the 1D steady state solution as described previously. The initial gas density is set to $n_{\rm hot} = 1.6 \times 10^{-4}\,$cm$^{-3}$ and $n_{\rm cold} = 1.6 \times 10^{-2}\,$cm$^{-3}$ in the hot and cold gases respectively. We use a resolution of $384\times128\times128$ in the box, which corresponds to a cell length of $0.78$~pc. This is approximately the minimum Field length in the simulation when thermal conduction is included. We also introduce a shear velocity profile across the front that takes the following form:
\begin{align}
    v_y = \frac{v_{\rm shear}}{2}\tanh(\frac{x}{a}),
\end{align}
where we set the scale length $a=5$~pc, and the shear velocity $v_{\rm shear} = 100$~km/s, which is of order the sound speed of the hot medium. The profile is then perturbed as follows to induce the Kelvin Helmholtz instability:
\begin{align}
    \delta v_x = A\exp(-\frac{x^2}{a^2})\sin(k_y y)\sin(k_z z),
\end{align}
where we set the perturbation amplitude $A$ to be $1\%$ of $v_{\rm shear}$. We also set the perturbation wavelength $\lambda_{i} = 2 \pi/k_{i}$ to be of order the box size, and set the ballistic speed of free electrons to be $V_m \sim 15$ times the hot gas sound speed when thermal conduction is included. The latter pertains only to \S\ref{sect:3dthermalconduction} -- thermal conduction is not included in any of the other 3D simulations. We check that results are not sensitive to these choices. Unlike adiabatic mixing layers which continue to grow over time, our mixing layers appear stable after the initial onset and development of turbulence. All quantities presented were measured in the latter half of the simulations after the mixing layers had been given sufficient time to reach this stage. The exact time periods vary between simulations, but simulations were run sufficiently long to ensure that they span at least 20 Myr. Errors bars reflect the standard deviation of the measured values. While the surface brightnesses $Q$ were saved at very small time intervals, and hence have many measurements, the turbulent velocities $u'$ were calculated from full simulation snapshots and have a smaller ($\sim 10$) number of measurements per simulation.

\subsection{Morphology of Mixing Layers: Transition from Single Phase to Multiphase}  \label{sect:morphology} 
\begin{figure*}
    \centering
    \includegraphics[width=\linewidth]{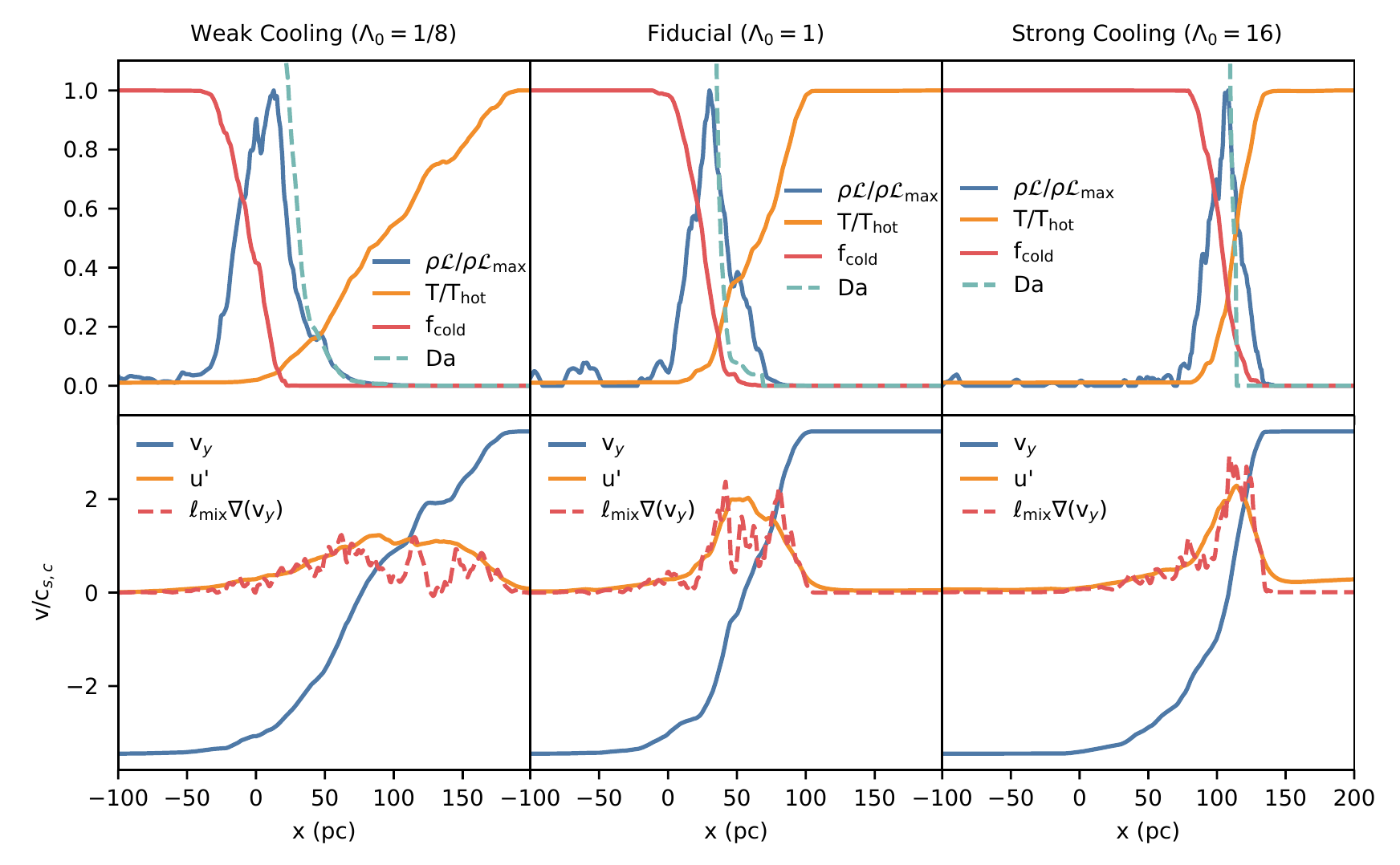}
    \caption{{\it Upper panels:} Normalized profiles for the mean emissivity, the mean temperature, and volumetric fraction of gas in the cold phase for various cooling strengths. The profile of the \Da number is also shown, denoting the region where mixing is more efficient than cooling. {\it Lower panels:} Mean shear and rms velocity profiles for the same selection of cooling strengths. The red dashed line shows the gradient of the shear velocity $\nabla(v_y)$ multiplied by a mixing length $\ell_{mix}=15$~pc, which traces the rms velocity profile well.}
    \label{fig:phase_profiles}
\end{figure*}

We begin by examining the morphology and slice averaged properties of the mixing layer, and how these vary with cooling (or equivalently, with \Da number $\mathrm{Da}$). We shall soon see (\S\ref{sect:mixing_length}) that temperature and velocity profiles can be calculated by judicious application of mixing length theory. 

The upper panels in Fig.~\ref{fig:phase_profiles} show the normalized profiles for the emissivity, mean temperature, and volumetric fraction of gas in the cold phase\footnote[4]{Defined to be $T < 5 \times 10^{4}$K gas.}, for the weak ($\Lambda_0 = 1/4$), fiducial ($\Lambda_0 = 1$), and strong ($\Lambda_0 = 8$) cooling cases respectively. They also plot the \Da profile. In calculating the \Da number, we use a fixed length scale $L=100$~pc (the box size in the direction of the flow), but use local values of the turbulent velocity $u^{\prime}$ and cooling time $t_{\rm cool}$ measured from the simulation. The initial \Da number ${\rm Da} = L/(u^{\prime} t_{\rm cool})$ in the hot medium is small due to the extremely long cooling times. However, as mixing proceeds and the mean temperature falls in the mixing layer, the cooling time falls and the \Da number {rises} toward cooler regions. The fact that the \Da number is a function of position within the mixing layer is important for understanding some key properties. Note that the mixing layer has roughly constant pressure. There are small pressure fluctuations seeded by cooling which are compensated  by increased turbulent pressure support, so that $P + \rho u^{\prime 2} \approx$~const (see Figure 8 of \citealp{ji19}), but these fluctuations are sufficiently small ($\delta P/P < 10\%$) that isobaric cooling is a good approximation. 

The lower panels in Fig.~\ref{fig:phase_profiles} show the corresponding mean and rms velocity profiles. The rms velocity is calculated by first subtracting off the mean flow in both the $y$ (flow) and $x$ (normal to cold/hot interface) directions. We have explicitly checked that the velocity dispersion is roughly isotropic ($\sigma_{x}^{2}=\sigma_{y}^{2}=\sigma_{z}^{2}$), a sign of well-developed turbulence. While stronger shear flows do display more anisotropy, the difference stays within a factor of two. It is interesting such isotropy can arise, despite the strong anisotropy in mean flow. The velocity dispersion $\sigma_{z}^{2}$ is a particularly good indicator, since there is no mean flow in the $z$ direction. 

Figure~\ref{fig:phase_profiles} reveals a number of interesting properties: 
\begin{itemize} 
\item{The criterion ${\rm Da}=1$ roughly controls the transition from single phase to multiphase gas, when the cold gas fraction first becomes non-zero. In the weak and fiducial cooling cases, the mean temperature falls substantially in the single phase regime (the `well-stirred reactor', in the language of \S\ref{sect:combustion}), before the gas turns multiphase. Thus, the mean temperature profile and the cold gas profile do not track one another. A substantial amount of the cooling flux is emitted in the single-phase regime. However, for the strong cooling case, all cooling takes place in the multiphase regime. In this case, the mean temperature profile tracks the cold gas profile; $\bar{T} \approx f_{\rm cold} T_{\rm cold} + (1-f_{\rm cold}) T_{\rm hot}$.}
\item{The turbulent velocity tracks the shear, $u^{\prime} \propto \nabla v_{y}$. This is expected from mixing length theory, where $u^{\prime} \approx l \nabla v_{y}$, and $l$ is the mixing length. We discuss this further in \S\ref{sect:uprime_scalings}}. 
\item{The normalized emissivity has an approximately Gaussian profile, as one would expect if cooling balances the divergence of turbulent diffusion. A diffusive process will of course have a Gaussian profile. For instance, in a multiphase medium, the fractal hot/cold gas boundary (where most of the cooling takes place in a thin sheet) has a Gaussian distribution of displacements from the mid-point, as expected for a random walk. The emissivity tracks the cold gas fraction rather than the mean temperature profile, peaking at $f_{\rm cold} \approx 0.5$. This makes sense, since the surface area of the hot-cold interface (which dominates cooling) peaks when $f_{\rm cold}=0.5$. The emissivity profile becomes narrower in the strong cooling regime. Later, we shall see that the area under the blue curves $Q \propto t_{\rm cool}^{-1/2}, t_{\rm cool}^{-1/4}$ in the weak and strong cooling regimes respectively.}
\item{In the strong cooling regime, the cooling emissivity and turbulence track one another closely. Both peak at the same spatial location (where $f_{\rm cool} \approx 0.5$). This was predicted by Eq.~\eqref{eq:emissivity_eff}, where $\tilde{\epsilon} \propto u^{\prime 1/2}$. This is consistent with our model, where turbulent mixing regulates the fraction of gas available for cooling.}
\item{The mean temperature profile and mean velocity profile also track one another quite closely, corresponding to $\mathcal{M} \approx 1 $ in the shear layers (see also Figure~9 of \citealt{ji19}). Thus, for instance, cooling causes the shear profile to narrow in moving from weak to fiducial cooling. This makes sense since turbulent diffusion governs both momentum and thermal transport. Furthermore, in the strong cooling case, the cooling emissivity also tracks the shear profile: $\tilde{\epsilon} \propto u^{\prime 1/2} t_{\rm cool}^{-1/2} \propto (\nabla v_{y})^{1/2} t_{\rm cool}^{-1/2}$. This correspondence fails when sink/source terms in the energy equation which are not present in the momentum equation become dominant: (i) very strong cooling (see the low temperature portion of the strong cooling regime), or (ii) highly supersonic flow in the hot medium (not shown). In the latter case, shocks and turbulent dissipation heat the gas, and so the hot region remains hot even when significant cool gas is mixed in. These effects narrow the temperature profile relative to the velocity profile.}
\end{itemize} 

\begin{figure*}
    \centering
    \includegraphics[width=\linewidth]{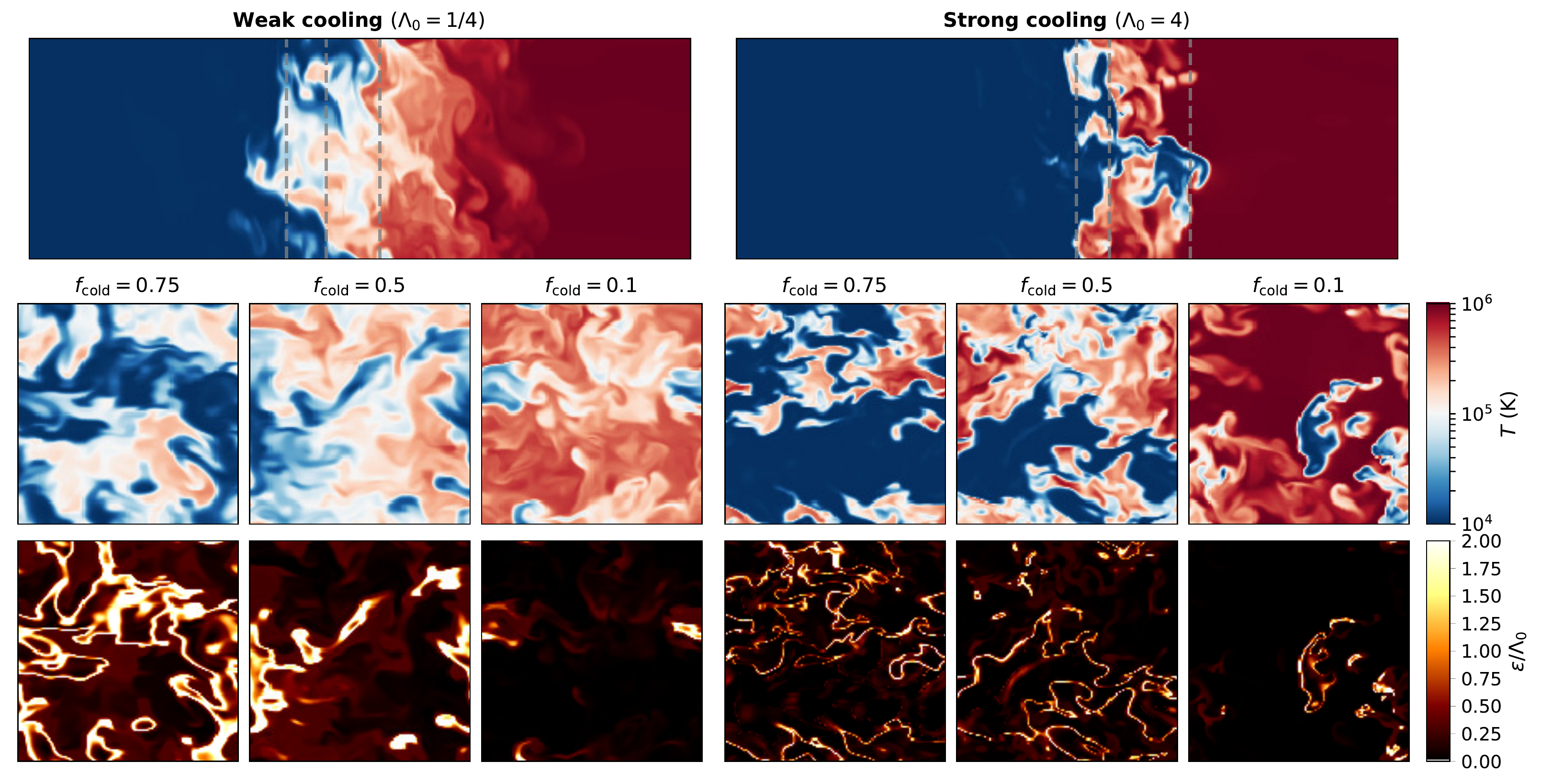}
    \hspace{-1cm}
    \caption{Slices of temperature and cooling for the `low cooling' (left) and `high cooling' regime where the cooling function has been reduced or amplified by a factor of four, respectively. The top row shows a temperature slice orthogonal to the flow while the middle and lower panels show temperature and emissivity at three different depths in the mixing layer (marked in the top with grey dashed lines). The cooling slices have been normalized by the boost factor of $\Lambda_0=1/4$ and $\Lambda_0 = 4$ for the left and right panels, respectively, to allow comparison of interface widths. We observe that the gas is strongly multiphase, especially when cooling is stronger, and that cooling happens mostly at the interfaces between the two phases. When cooling is weaker, these interfaces are thicker. This is consistent with the idea that they are defined by a diffusive length $\lambda_{\rm F} \propto t_{\rm cool}^{1/2}$.}
    \label{fig:snapshots}
\end{figure*}

The distinction between the weak (single phase) and strong (multiphase) cooling regimes can be clearly seen in Fig.~\ref{fig:snapshots}. At first glance, both cases appear to be similar, except that the weak cooling case has a broader mixing layer (top panel). However, it is already apparent that there is a lot more intermediate temperature ($T\sim 10^{5}$K) gas in the weak cooling case. We can also see this in the temperature slices, which only show the `multiphase' portion of both cases (when $f_{\rm cold}$ is non-zero). For weak cooling, the `hot' phase in this regime is significantly cooler than $T=10^{6}$K, the initial temperature of the hot gas -- it has cooled via efficient mixing in the single-phase regime. By comparison, the temperature contrast between the two phases is much higher in the strong cooling case, with a clearly bimodal temperature distribution centered at $T \sim 10^{4}\,$K and $T\sim 10^{6}\,$K. In both cases, the amount of intermediate temperature ($T\sim 10^{5}$K) gas peaks when $f_{\rm cool} \sim 0.5$, where the emissivity also peaks. In the strong cooling case, cooling is clearly dominated by the very thin (unresolved) interface between the phases, as can be seen in the emissivity slices (bottom panel). This is less true in the weak cooling case, where a larger fraction of the volume contributes to cooling (note the low filling factor of interface regions at $f_{\rm cold}=0.5$, when cooling peaks). Furthermore, the interface regions (which should scale as $l \propto t_{\rm cool}^{1/2}$) are now broader and clearly resolved.

\subsection{Scaling Relations} 

The key theoretical quantity of interest in radiative mixing layers is the hot gas entrainment rate $v_{\rm in}$, or equivalently the  surface brightness $Q$ (assuming that hot gas enthalpy flux balances cooling). This determines the rate  at which hot gas is converted to cold gas, which has many important implications, amongst them the ability of cold gas to survive in the face of hydrodynamic instabilities \citep{gronke18}. In previous work, we derived the mass growth rate shown in Eq.~\eqref{eq:vin}. However, this was performed at relatively low resolution. Higher resolution work similar to that done here \citep{ji19} also obtained scaling relations, with some important differences. However, their results relied on a rather small number of simulations. Here, we clarify the nature of the scaling relations using a larger set of simulations, and thus put the results obtained from previous studies in a broader context. In particular, we explicitly test\footnote[5]{In practice, we only vary $t_{\rm cool}$ when testing the scaling $Q \propto (L/t_{\rm cool})^{n}$. Since the cooling length is the only scale in the problem (ideal hydrodynamics is scale free), varying $L$ and $t_{\rm cool}$ at fixed $L/t_{\rm cool}$ are equivalent. We have checked this previously for cloud-crushing setups.}  the predicted scaling relations Eqs.~\eqref{eq:vin_weak_cooling} and \eqref{eq:fiducial-scaling}, which state that in the weak cooling regime, $Q \propto v_{\rm in} \propto u^{\prime 1/2} (L/t_{\rm cool})^{1/2}$, while in the strong cooling regime, $Q \propto v_{\rm in} \propto u^{\prime 3/4} (L/t_{\rm cool})^{1/4}$, with no additional dependence on other parameters such as overdensity $\chi$ and flow  Mach number (relative to the hot gas sound speed) $\mathcal{M}$ (\S\ref{sect:Q_scalings}). We then test scalings for emissivity, or equivalently for the width of the mixing layer (\S\ref{sect:emiss_scalings}). Finally, we test how turbulent velocities $u^{\prime}$ vary with $\chi,\mathcal{M},t_{\rm cool}$ in our specific setup (\S\ref{sect:uprime_scalings}).

\subsubsection{Scaling Relations for Q}
\label{sect:Q_scalings} 

\begin{figure}
    \centering
    \includegraphics[width=\columnwidth]{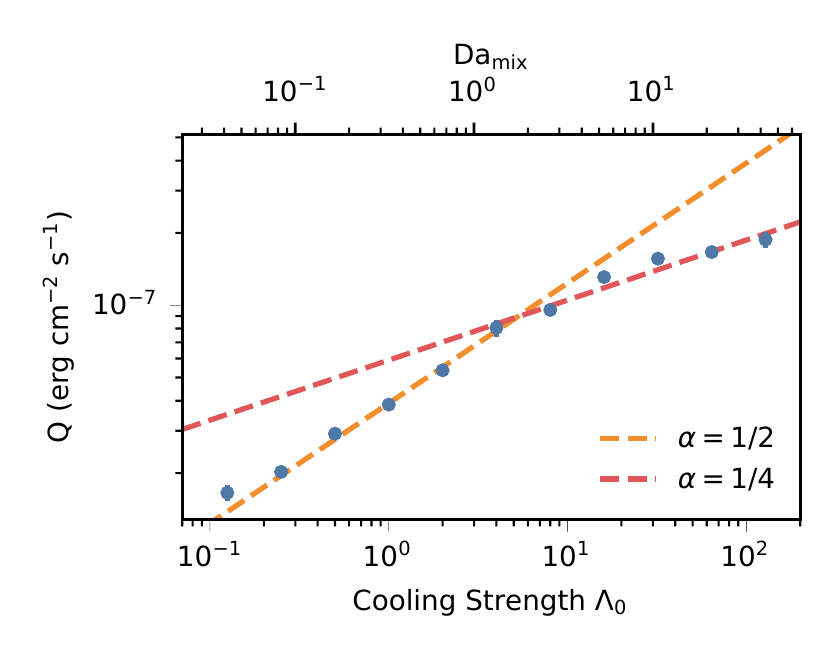}
    \caption{Surface brightness $Q$ as a function of cooling strength. 
    We see that $Q$ scales as $t_{\rm cool}^{-1/2}$ for weak cooling and $t_{\rm cool}^{-1/4}$ for strong cooling. These two regimes can be characterized by ${\rm Da}_{\rm mix}$, which we show in the top axis for reference.}
    \label{fig:coolvscool}
\end{figure}

\begin{figure}
    \centering
    \includegraphics[width=\columnwidth]{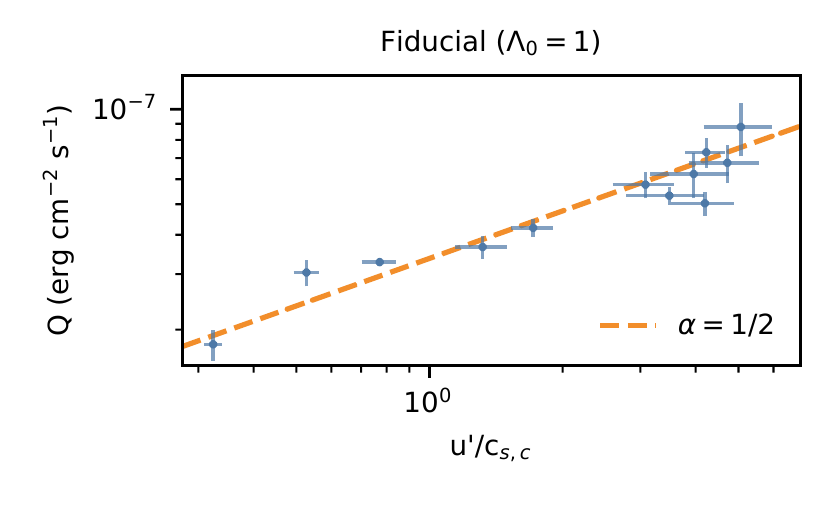}
    \includegraphics[width=\columnwidth]{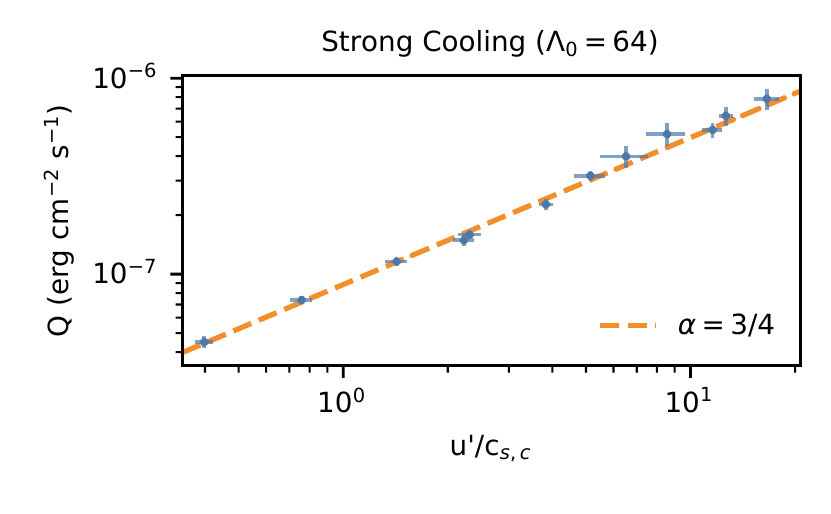}
    \caption{Surface brightness $Q$ plotted against turbulent velocities for the two cooling regimes. Expected scalings are given by the orange dashed lines.}
    \label{fig:shear_dependence}
\end{figure}

\begin{figure} 
    \centering
    \includegraphics{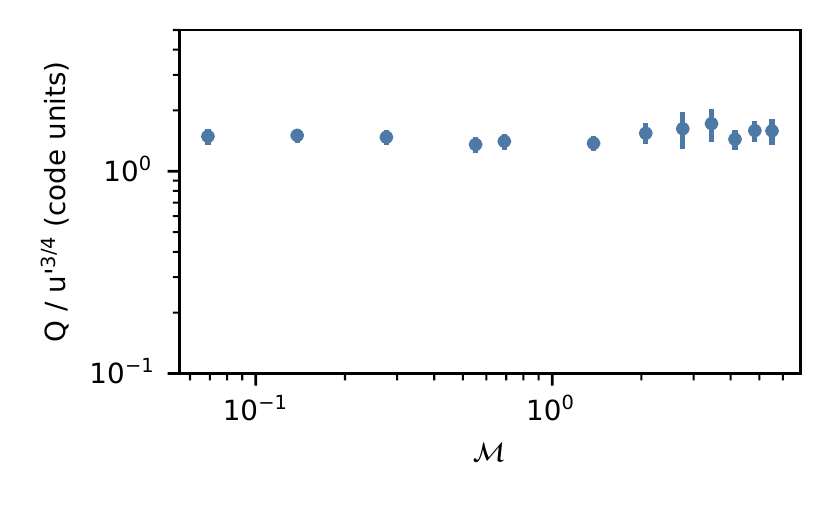} 
    \includegraphics{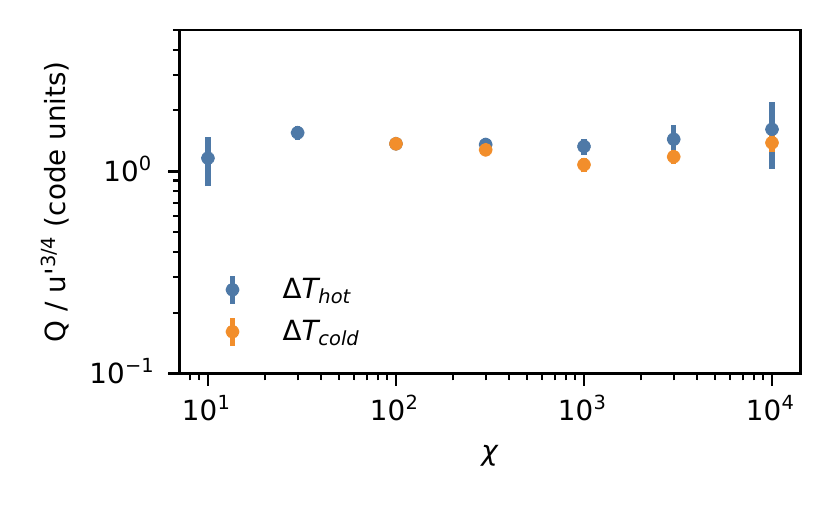}
    \caption{$Q/u'^{3/4}$ is independent of both the shear velocity (upper panel) and the overdensity (lower panel) as expected from Eq.~\eqref{eq:fiducial-scaling}. Note that to change the overdensity we varied both the cold and hot gas temperatures (shown in blue and orange, respectively).}
    \label{fig:money_plot} 
\end{figure} 

{\it Dependence on Cooling.} Figure~\ref{fig:coolvscool} shows the surface brightness $Q$ as a function of cooling strength $\Lambda_0$. It is clear that $Q \propto \Lambda^{1/2}\propto t_{\rm cool}^{-1/2}$ ($Q\propto\Lambda^{1/4}\propto t_{\rm cool}^{-1/4}$) in the weak (strong) cooling regimes, as predicted by Eqs.~\eqref{eq:vin_weak_cooling} and \eqref{eq:fiducial-scaling} respectively. We have already seen that the \Da number varies spatially across a mixing layer. Here, it is useful to define a \Da number characterizing a single simulation as a whole. This provides a reference point for differentiating between the two regimes. Thus, for each simulation, we have to choose a single value for the spatially varying quantities $u^{\prime},t_{\rm cool}$. We choose the peak value of $u^{\prime}$; later, we shall see in \S\ref{sect:uprime_scalings} that this is insensitive to cooling. For $t_{\rm cool}$, we use the cooling time of mixed intermediate $T=2\times10^5$~K gas, which is $t_{\rm cool,mix}=10$~Myr for the cooling time in the fiducial simulation, and adjust it accordingly in other simulations. We denote the resulting characteristic \Da number as ${\rm Da}_{\rm mix}$, which is shown in the top of Fig.~\ref{fig:coolvscool} as a secondary axis. The turnover between the two scalings thus occurs where ${\rm Da}_{\rm mix} \sim 1$. When ${\rm Da}_{\rm mix}$ is small, we are in the weak cooling regime and conversely, when ${\rm Da}_{\rm mix}$ is large, we are in the strong cooling regime.

{\it Dependence on Turbulence.} Figure~\ref{fig:shear_dependence} shows the surface brightness $Q$ as a function of the measured peak turbulent velocity $u^{\prime}$, in the weak and strong cooling regimes respectively. The turbulent velocity was varied by changing the shear velocity ($u^{\prime} \propto v_{\rm shear}^{0.8}$; see \S\ref{sect:uprime_scalings}). As given by Eqs.~\eqref{eq:vin_weak_cooling} and \eqref{eq:fiducial-scaling}, $Q\propto u^{\prime 1/2}$ and $Q\propto u^{\prime 3/4}$ in the weak and strong cooling regimes, respectively. Interestingly, these relationships stay the same even when the flow is supersonic with respect to the hot gas.

{\it No Hidden Parameters.} As discussed in \S\ref{sect:combustion}, radiative mixing layers are characterized by the dimensionless parameters ${\rm Da}=\tau_{\rm turb}/t_{\rm cool}$, $\mathcal{M}, \chi$. Above, we tested the dependence on Da via the dependence on $\tau_{\rm turb}$ and $t_{\rm cool}$. By contrast, our theoretical predictions for $Q$ (Eqs.~\eqref{eq:vin_weak_cooling} and \eqref{eq:fiducial-scaling}) do not contain any explicit dependence\footnote[6]{$Q$ does, however, have {\it implicit} dependence on $\mathcal{M}, \chi$ since $u^{\prime}=u^{\prime}(\mathcal{M},\chi)$. See \S\ref{sect:uprime_scalings}} on $\mathcal{M}, \chi$. We confirm this by plotting $Q$ divided by our fiducial scalings against $\chi$ and $\mathcal{M}$ in Fig.~\ref{fig:money_plot}. For the simulations that vary the overdensity, we assumed a flat cooling curve and adjusted $\Lambda_0$ to keep $t_{\rm cool}$ of the cold gas constant throughout. We see that our fiducial scalings are accurate, with no additional dependence on the parameters  $\chi$ and $\mathcal{M}$ across a wide dynamic range. While Fig.~\ref{fig:money_plot} shows the strong cooling case ($\Lambda_0=64$), we also check that this holds for the scaling in the fiducial regime.

In summary, our fiducial formula for $Q$ is
\begin{align}
    Q = Q_{0}
    \left( \frac{P}{160 k_{\rm B} {\rm cm^{-3}~K}} \right)
    \left( \frac{u'}{30{\rm~km/s}} \right)^{1/2}
    \left(\frac{L}{100~{\rm pc}}\right)^{1/2}
    \left(\frac{t_{\rm cool,min}}{0.03{\rm ~Myr}}\right)^{-1/2}
    \label{eqn:Q_slow}
\end{align}
in the slow cooling regime and
\begin{align}
    Q = Q_{0}
    \left( \frac{P}{160 k_{\rm B} {\rm cm^{-3}~K}} \right)
    \left( \frac{u'}{30{\rm~km/s}} \right)^{3/4}
    \left(\frac{L}{100~{\rm pc}}\right)^{1/4}
    \left(\frac{t_{\rm cool,min}}{0.03{\rm ~Myr}}\right)^{-1/4}
    \label{eqn:Q_fast}
\end{align}
in the fast cooling regime, where $t_{\rm cool}$ is evaluated at the peak of the cooling curve and the scalings are normalized by $Q$ at the turnover point in Fig.~\ref{fig:coolvscool}:
\begin{align}
    Q_{0} \sim 8.8 \times 10^{-8}{\rm ~erg}{\rm ~cm}^{-2}{\rm ~s}^{-1}.
\end{align}

\subsubsection{Scaling relations for Effective Emissivity}
\label{sect:emiss_scalings} 

\begin{figure}
    \centering
    \includegraphics{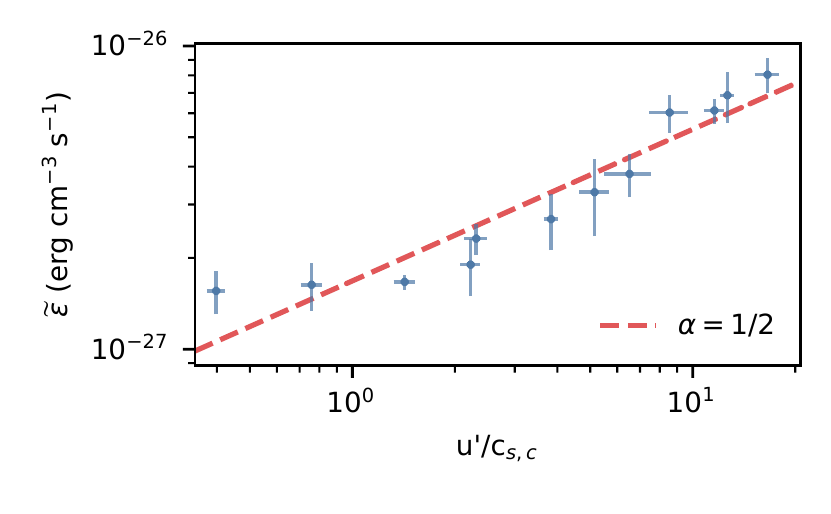}
    \includegraphics{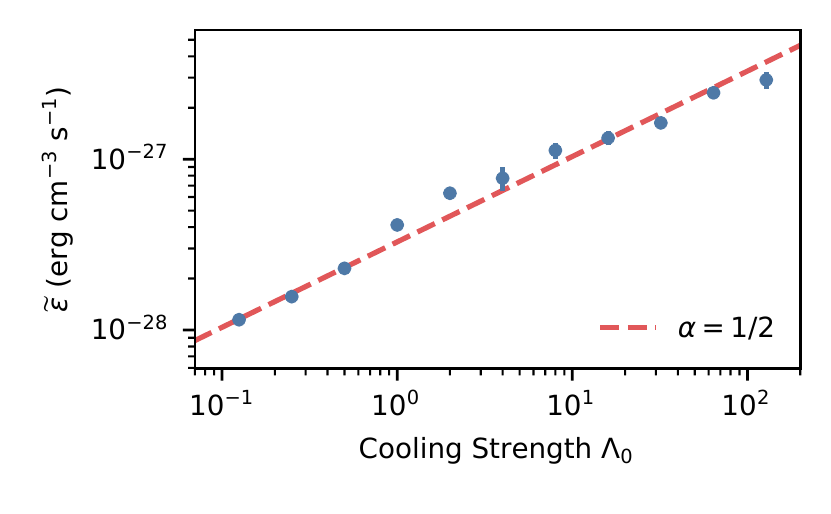}
    \caption{The effective emissivity for various shear velocities with $\Lambda_0=64$ and for different cooling strengths. Expected scalings are given by the red dashed lines.}
    \label{fig:emissivity_scalings} 
\end{figure}

The above simulations confirm our scalings for $Q$. However, we would like to test the theoretical ideas behind them. The formula in the single-phase regime (Eq.~\eqref{eq:vin_weak_cooling}) is a straightforward application of mixing length theory, entirely analogous to the thermal conduction case (\S\ref{sect:1D}). However, the formula for the multiphase regime,  (Eq.~\eqref{eq:fiducial-scaling}) is much less well-established. Its central claim is that the eddy lifetime is now the geometric mean of the eddy turnover time and the cooling lifetime (Eq.~\eqref{eq:tilde_tau}). As discussed in \S\ref{sect:fiducial_scaling}, this can be tested by checking that the effective emissivity of the multiphase medium scales as $\tilde{\epsilon} \propto u^{\prime 1/2} t_{\rm cool}^{-1/2}$ (Eq.~\eqref{eq:emissivity_eff}), and that the width of the multiphase region scales as $h \propto {\rm Da}^{-1/4} \propto u^{\prime 1/4} t_{\rm cool}^{1/4}$. Note that these two quantities are related by $Q \propto \tilde{\epsilon} h$, so only one of them constitutes an independent test. In Fig.~\ref{fig:emissivity_scalings}, we show $\tilde{\epsilon}$ at the spatial location where it is maximized, as a function of $u^{\prime}$ and cooling strength $\Lambda \propto t_{\rm cool}^{-1}$. It clearly conforms to the expected scalings. We also confirm that the FWHM of the multiphase region agrees with the predicted scalings, though this is a less well-defined and noisier quantity. In \S\ref{sect:mixing_length}, we shall see that the predicted effective emissivity allows remarkably accurate predictions of mean temperature profiles. 

In summary, our fiducial formula for effective emissivity in the strong cooling regime is the Gaussian model
\begin{equation}
    \tilde{\epsilon} = c_p \frac{P}{\tilde{\tau}_{\rm cool}} \mathcal{N}(0,\sigma^{2}),
\label{eq:true_emissivity} 
\end{equation}
where from Eq.~\eqref{eq:tilde_tau}:
\begin{align}
    \tilde{\tau}_{\rm cool} =
    2.5 \, {\rm Myr} \left( \frac{L}{100 \, {\rm pc}} \right)^{1/2} \left( \frac{u'}{30{\rm~km/s}} \right)^{-1/2}
        \left(\frac{t_{\rm cool,min}}{0.03{\rm ~Myr}}\right)^{1/2},
    \label{eq:model_cooling} 
\end{align}
where $u^{\prime}$ is evaluated at its peak (and follows the scaling relations obtained in \S\ref{sect:uprime_scalings}), and $t_{\rm cool}$ is evaluated at the peak of the cooling curve. This is the form plotted as the red dashed lines in Fig.~\ref{fig:emissivity_scalings}; note that it is a derived quantity with no free parameters. From Eqs.~\eqref{eqn:Q_fast}, \eqref{eq:true_emissivity} and \eqref{eq:model_cooling}, we obtain the width of the cooling region as
\begin{align}
    \sigma \sim 16 {\rm ~pc~}
    \left( \frac{L}{100 \, {\rm pc}} \right)^{3/4} 
    \left( \frac{u'}{30{\rm~km/s}} \right)^{1/4}
    \left(\frac{t_{\rm cool,min}}{0.03{\rm ~Myr}}\right)^{1/4}.
    \label{eqn:emissivity_width}
\end{align}
In the weak cooling (homogeneous reactor) regime, the emissivity is unchanged from the standard $\epsilon=n^{2} \Lambda(T)$ form.

\subsubsection{Scaling Relations for $u^{\prime}$}
\label{sect:uprime_scalings} 

\begin{figure}
    \centering
    \includegraphics{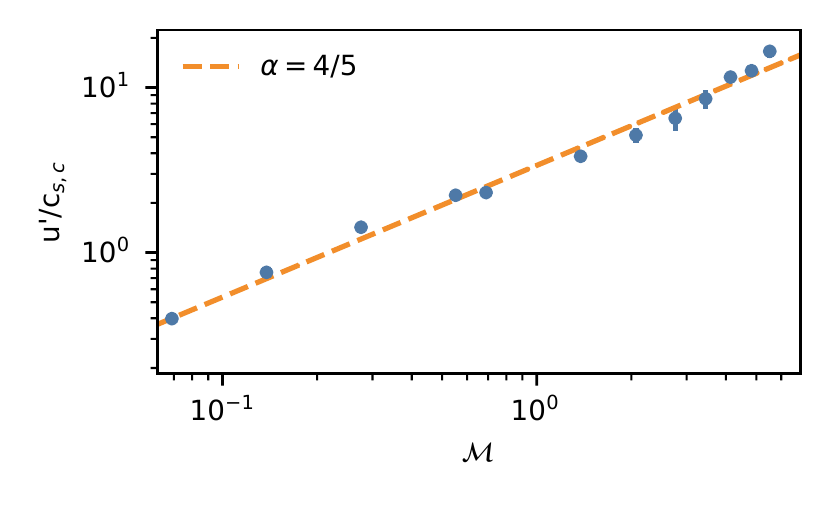}
    \includegraphics{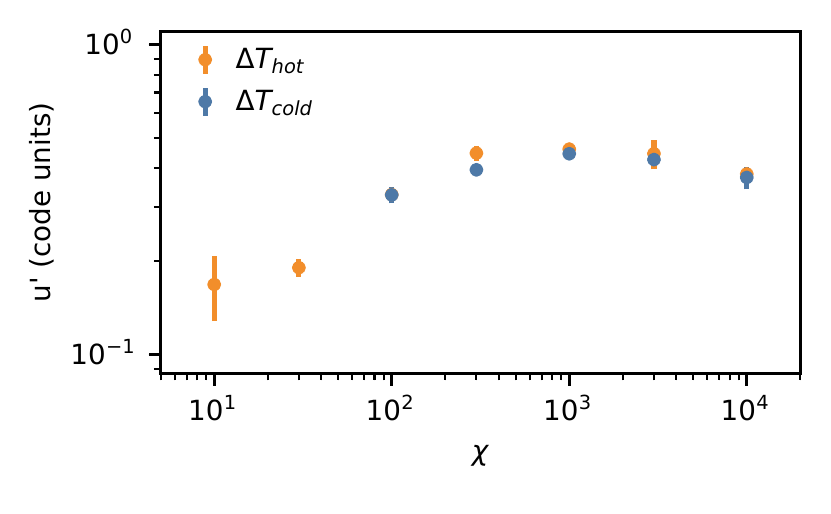}
    \includegraphics{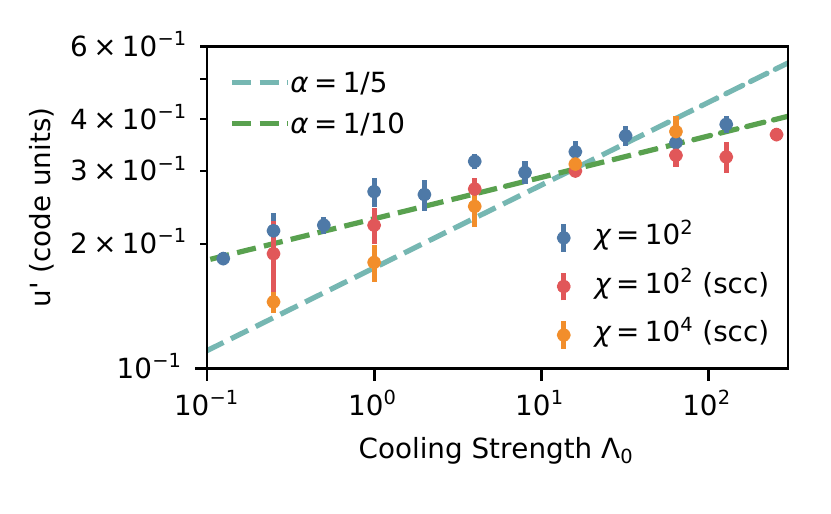}
    \caption{Dependence of turbulent velocity on shear velocity (using $\chi\sim 100$), overdensity and cooling. Overdensity and shear velocity are varied in the strong cooling regime ($\Lambda_0 = 64$). Scalings are represented by dashed lines. The middle panel and bottom panel includes simulations with a simplified cooling curve (see text).}
    \label{fig:uprime_scalings}
\end{figure}

We now consider how the turbulent velocity $u^{\prime}$ depends on other parameters in our simulation, specifically cooling, the overdensity $\chi$ and shear Mach number $\mathcal{M}$. We focus on the strong cooling regime, since that is of the most astrophysical interest (e.g., for cloud survival, \citealt{gronke19}) and less well-understood. We stress that these scalings are particular to our setup and not as general as the scalings for $Q$. They will differ depending on the source of turbulent driving. Thus, we do not invest the same effort in deriving and understanding them. 

It is still useful to note some theoretical considerations. The free energy for driving turbulence in our mixing layer is shear, where: 
\begin{equation}
    u^{\prime 2} \approx l^{2} \left( \frac{\partial v_{y}}{\partial x} \right)^{2},
\label{eq:vorticity} 
\end{equation}
with $l$ being the characteristic size of vortices.
This is simply the statement that the vorticity of the eddies and the shear flow (from which the eddies derive their vorticity) are comparable. Since we volume average within slices through the mixing layer when measuring $u^{\prime}$, the contribution is dominated by the turbulent velocity of the hot gas in the region where turbulence $u^{\prime}$ and emissivity peak ($f_{\rm hot} \sim 0.5$). In our simulations, we indeed see $u^{\prime} \propto \grad v_y$ (Fig.~\ref{fig:phase_profiles}). Cooling can play an important role in regulating the width of the mixing layer, and hence $\grad v$ and $u^{\prime}$ (see Fig.~\ref{fig:phase_profiles}). However, once deep in the multiphase regime, cooling only has a weak effect. We can see this from $\grad v \sim v_{\rm shear}/h$; if we combine our prediction for $h$, Eq.~\eqref{eq:h_fast_cooling}, and Eq.~\eqref{eq:vorticity}, we obtain:
\begin{equation} 
    u^{\prime} \propto v_{\rm shear}^{4/5} \left( \frac{L}{t_{\rm cool}} \right)^{1/5}.
    \label{eq:uprime_scalings}
\end{equation}
Note that for a given shear velocity and cooling time, there is no explicit overdensity dependence. In the multi-phase Da$\gg 1$ regime, $t_{\rm cool}$ is the cooling time of the cold gas.

Figure~\ref{fig:uprime_scalings} show the scalings of $u^{\prime}$ with $v_{\rm shear}$, $\chi$, and $\Lambda_{0}$. The upper panel includes the scalings from previously mentioned simulations that vary the shear velocity from the fiducial setup. We find, as in Eq.~\eqref{eq:uprime_scalings} that $u^{\prime} \propto v_{\rm shear}^{4/5}$.

Changing the overdensity involves a change in the temperature regime where cooling takes place. We want to do so while keeping $t_{\rm cool}$ constant. We adopt a simplified form of the cooling curve $\Lambda$~=~const for $T < 10^{4}$K and $\Lambda \propto T^4$ for $T > 10^4$~K, such that $t_{\rm cool} \propto T^{-2}$ and $t_{\rm cool} \propto T^{2}$ below and above $T = 10^{4}$ K respectively. This singles out the minimum in the cooling time to occur at $T \sim 10^{4}$K (as is true of more realistic cooling curves). This cooling time is held constant. To vary $\chi$, we keep either the hot or cold phase constant and change the temperature and density of the other phase.

In the middle panel of Fig.~\ref{fig:uprime_scalings}, we see that for $\chi \gtrsim 100$, $u^{\prime}$ behaves according to Eq.~\eqref{eq:uprime_scalings} with no dependence on $\chi$, regardless of whether the hot or cold phase is being varied. On the other hand, at lower overdensities, $u^{\prime}$ does not scale as expected: it declines toward low $\chi$ instead. In this low overdensity regime, the temperature and velocity profiles decouple and no longer track one another. This was first noted by \citet{fielding20} (see their Figure 1), but we draw a slightly different conclusion from them: the decoupling of thermal and momentum profiles is not general but only happens at low $\chi$. The reason is that the cooling time of mixed gas ($t_{\rm cool,mix} \sim \chi t_{\rm cool,cold}$) is still relatively short, where $T_{\rm mix} \sim (T_{\rm hot} T_{\rm cold})^{1/2}$. The hot gas then rapidly cools after a small amount of mixing with cold gas. Radiative cooling outpaces momentum transport, which mostly takes place when the gas is already cold; the velocity shear and turbulence peak in the single phase regime. This vitiates the assumptions behind Eq.~\eqref{eq:uprime_scalings}. Indeed, the assumption of a thin mixing layer is no longer valid. For the ratio of the thickness of the shear layer $h$ and the the horizontal length  $L_{y}$ we can write to first order
\begin{equation}
    \frac{h}{L_{y}} \sim \frac{v_{\rm in}}{v_{\rm shear}} \sim \frac{c_{\rm s,cold}}{{\mathcal M} c_{\rm s,hot}} \sim \frac{1}{\mathcal{M} {\sqrt{\chi}}},
\end{equation}
where the first equality comes from the continuity equation. Hence, the flow decelerates on a length scale comparable to the thickness of the mixing layer as the simulation proceeds. Overall, this regime holds less astrophysical significance: because of the location of the stable phases in the cooling curve, most situations of astrophysical interest involve density contrasts $\chi \gtrsim 100$, where Eq.~\eqref{eq:uprime_scalings} holds. 

In the bottom panel of Fig.~\ref{fig:uprime_scalings}, we check the dependence of $u^{\prime}$ on $t_{\rm cool}$. The $\chi=10^4$ simulation follows the expected $u^{\prime} \propto \Lambda^{0.2}$ scaling. However, the $\chi=100$ simulation follows a slightly weaker $u^{\prime} \propto \Lambda^{0.1}$ scaling. This remains true for simulations which use the full (realistic) cooling curve (blue points). This is because the turbulent velocity approaches equation \eqref{eq:uprime_scalings}  asymptotically as $\chi$ increases. For instance, it only becomes fully independent of overdensity for $\chi \gtrsim 300$ (middle panel). In any case, the difference is small.

\subsection{Comparing Simulations to 1D Mixing Length Models} 
\label{sect:mixing_length} 

\begin{figure}
    \centering
    \includegraphics[width=\columnwidth]{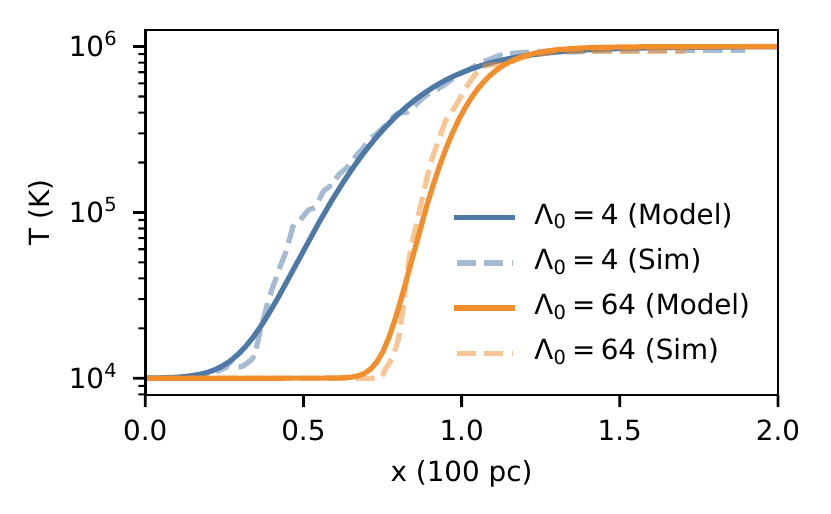}
    \caption{A comparison between simulation profiles and a 1D mixing length          model at two different cooling strengths shows good agreement.}
    \label{fig:mixing_length}
\end{figure}

\begin{figure}
    \centering
    \includegraphics[width=\columnwidth]{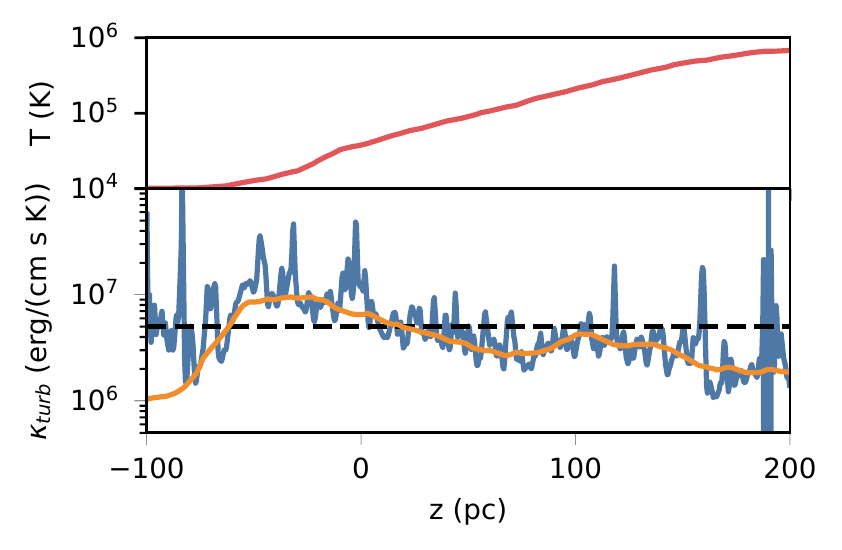}
    \caption{Measurement of turbulent diffusion through the mixing layer in a simulation without radiative cooling. The upper panel shows the corresponding average temperature. In the lower panel, the blue line shows $\kappa_{\rm turb}$ measured from the simulation via Eq.~\eqref{eq:kappa_turb}, while the orange line shows the mixing length approximation for the fiducial setup (cf. \S\ref{sect:mixing_length}). The dashed line shows the value used for the profiles shown in Fig.~\ref{fig:mixing_length}}.
    \label{fig:measureturb}
\end{figure}

\begin{figure} 
    \centering
    \includegraphics[width=\columnwidth]{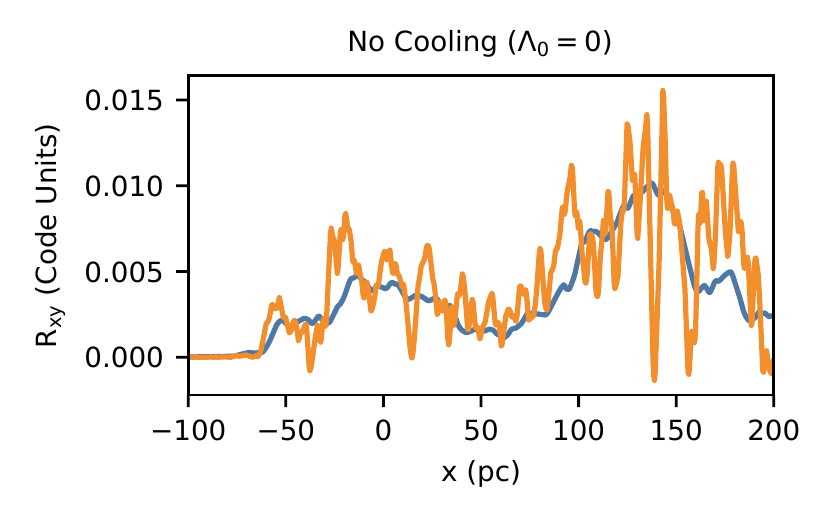}
    \includegraphics[width=\columnwidth]{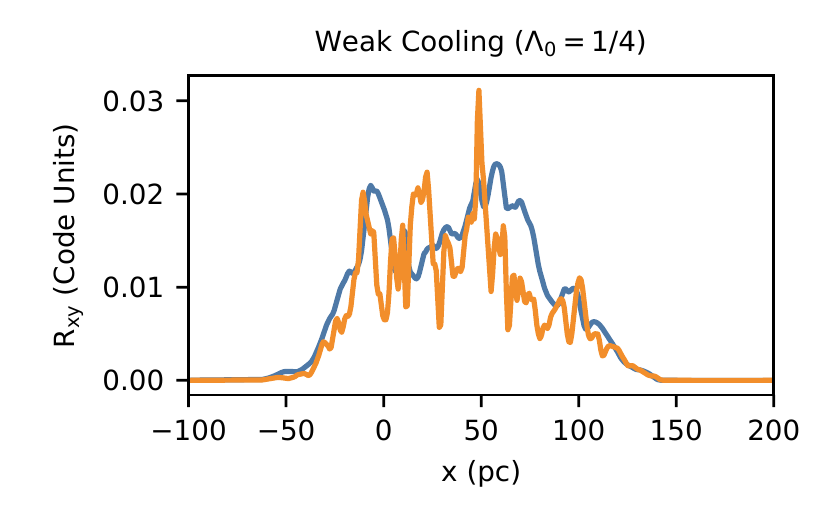}
    \includegraphics[width=\columnwidth]{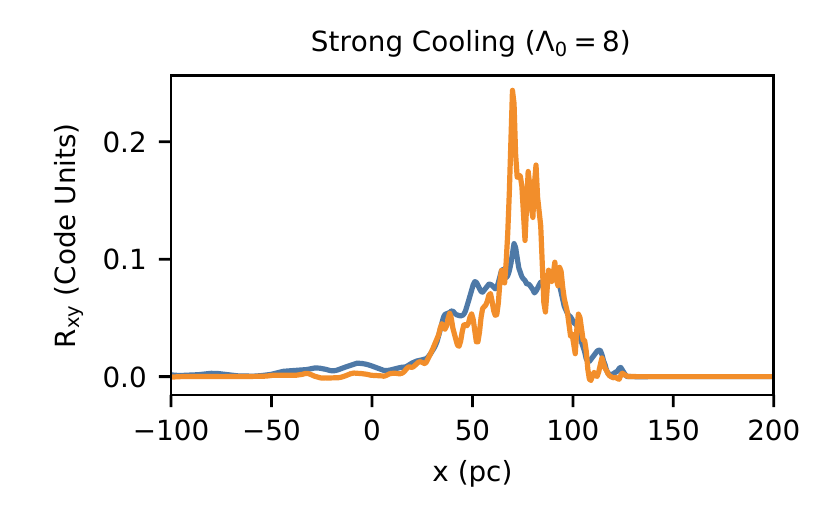}
    \caption{For adiabatic, weak and strong cooling, the Reynolds stress is shown in blue along with the mixing length model in orange. As in the case of turbulent heat transport, we find good agreement between them.}
    \label{fig:Reynolds_stress}
\end{figure} 

In \S\ref{sect:1D}, we constructed time-steady 1D models of thermal interfaces, where thermal conduction balances radiative cooling. In the single-phase, weak cooling case, one can construct similar profiles by substituting turbulent heat diffusion for thermal conduction (as has been done for galaxy clusters, \citealp{kim03,dennis05}). However, it may seem absurd to carry this over to the strong cooling regime, where the highly fluctuating, fractal and multiphase structure of the radiative front seems to preclude a simple mean-field approach. Here, we show that with judicious choice of the effective emissivity, such a model matches simulations surprisingly well. 

We first establish that in an isobaric medium, we can write the turbulent heat flux in a form similar to that for thermal conduction, $F_{\rm turb} = -\kappa_{\rm turb} \nabla T$. In mixing length theory, the turbulent heat flux is proportional to the gradient of specific entropy (e.g., \citealt{dennis05}): 
\begin{equation}
    F_{\rm turb} = -D_{\rm eddy} \rho T \nabla s, 
\end{equation}
where $s=c_{\rm V} {\rm ln}(p/\rho^{\gamma})$ is the specific entropy, $\gamma=c_{\rm P}/c_{\rm V}=5/3$ is the ratio of specific heats, $c_{\rm V}= 3/2 k_{\rm B}/\bar{m}$ is the specific heat at constant volume, and $D_{\rm eddy}$ is the eddy diffusivity, with units $[D_{\rm eddy}]=L^{2} T^{-1}$. However, under isobaric conditions $\nabla P=0$, evaluating the above expression gives: 
\begin{equation}
    F_{\rm turb} = -D_{\rm eddy} \rho c_{\rm P} \nabla T,
\end{equation}
i.e., the turbulent heat flux is proportional to the temperature gradient. In an isobaric medium, one can equally well think of passive scalar advection of entropy or temperature\footnote[7]{Of course, only entropy advection is correct in a stratified medium like a star or galaxy cluster, where mixing length theory is usually applied.}. For simplicity, and analogous to thermal conduction, we will consider $F_{\rm turb} = -\kappa_{\rm turb} \nabla T$, where: 
\begin{equation}
    \kappa_{\rm turb} = D_{\rm eddy} \rho c_{\rm P}. 
\end{equation}

We then assume that the coefficient $\kappa_{\rm turb}$ is a constant independent of temperature. We can show after the fact that this is a reasonable assumption. As in \S\ref{sect:1D}, we then solve the 1D steady-state equation:
\begin{equation} \label{eqn:newODE}
   \kappa_{\rm turb} \frac{\dd^2 T}{\dd x^2} = j_x c_p \frac{\dd T}{\dd x} + \rho \mathcal{L}.
\end{equation}
In \S\ref{sect:1D}, the thermal conductivity $\kappa$ was known and we solved for the mass flux $j_{\rm x} = \rho v$ as an eigenvalue. Here, since the medium is multiphase, the emissivity is not the same as that of single-phase medium with the same mean temperature. Motivated by the scalings in \S\ref{sect:emiss_scalings}, we model the emissivity as a Gaussian as specified in Eq.~\eqref{eq:true_emissivity}. This sets $Q$ and hence the mass flux $j_{\rm x}= Q/c_{\rm P} (T_{\rm hot}-T_{\rm cold})$. Because our cooling is now a function of position and not temperature, we specify the value of $\kappa_{\rm turb}$ and solve for the profile via the shooting method, subject to the same boundary conditions as before (Eq.~\eqref{eq:BCs}). The results for two strong cooling ($\Lambda_0 = 4,64$) cases are shown in Fig.~\ref{fig:mixing_length}, where we have used $\kappa_{\rm turb} = 5 \times 10^{6} \, {\rm erg \, cm^{-1} \, s^{-1} \, K^{-1}}$ for both cases. This is justified below. Such a simple mixing length model provides a remarkably good fit to the profiles seen in our simulations. This allows the construction of rapid semi-analytic models of radiative mixing layers, which is very useful when comparing against observations (e.g., matching line column density ratios) when the underlying model parameters such as $u^{\prime},L, t_{\rm cool}$ are unknown and one has to search over parameter space. We can thus construct models for multiphase mixing layers with the same speed and ease as for thermal conduction.  

We estimate the turbulent diffusion coefficient by applying mixing length theory to direct measurements from the simulation. The simplest way to do this is to first obtain $\kappa_{\rm turb}$ from adiabatic simulations where we can model the turbulence using the Reynold's Averaged Naiver-Stokes (RANS) equations. This approach separates the flow into two components: a mean flow and a time-dependent varying flow. By representing initial variables such as temperature and velocity as $\phi = \bar{\phi} + \phi''$ where $\bar{\phi} $ is the density weighted mean value, we find an extra term of the form $\langle \rho v''_x T'' \rangle$ which is modelled using a simple gradient relation:
\begin{equation}
    F_{\rm turb} = \langle c_p \rho v''_x T'' \rangle = -\kappa_{\rm turb}\frac{\dd\bar{T}}{\dd x}.
    \label{eq:kappa_turb} 
\end{equation}
Figure \ref{fig:measureturb} shows the resulting measurement of $\kappa_{\rm turb}$ as a function of height in the mixing layer for a single time snapshot where the averages were taken along the $y$-$z$ plane. Consistent with our assumptions, $\kappa_{\rm turb}$ is roughly constant, and the dashed line shows the value we adopt in the simple model above. The solid orange line in Fig.~\ref{fig:measureturb} shows the mixing length approximation: 
\begin{equation}
    \kappa_{\rm turb} =\bar{\rho} c_{\rm P} l u^{\prime} ,
    \label{eq:kappa_mix}
\end{equation}
with a mixing length of $l=4$~pc, which fits the result of Eq.~\eqref{eq:kappa_turb} from the simulation remarkably well.

In mixing length theory, the mixing length $l$ cannot be obtained from first principles, but must be calibrated from experiments or simulations. Nonetheless, the value we obtain is reasonable from order of magnitude considerations. Since $u^\prime \approx l \nabla v_{\rm y} \sim (l/L)v_{\rm shear}$, we have: 
\begin{equation}
    l \sim \frac{u^{\prime}}{v_{\rm shear}} L \sim \frac{c_{\rm s,cold}}{c_{\rm s,hot}} L \sim \frac{L}{\sqrt{\chi}},
\end{equation}
which gives $l \sim 10$~pc for our setup.

Due to the multiphase structure of the mixing layers with strong cooling, it is not possible to use Eq.~\eqref{eq:kappa_turb} to measure $\kappa_{\rm turb}$ there. It is interesting that $\kappa_{\rm turb}$ derived from adiabatic simulations provides a good fit when used to solve for temperature profiles in strong cooling simulations, and is consistent with the finding that cooling appears to have little effect on turbulence. 

Instead, in cooling simulations we can focus on velocity profiles to verify the mixing length approach. In Fig.~\ref{fig:Reynolds_stress}, we plot the Reynolds stress in adiabatic, weak, and strong cooling simulations, and compare with the expectation from mixing length theory that
\begin{equation}
    R_{\rm xy} = -\langle \rho v_{x}^{\prime\prime} v_{y}^{\prime\prime} \rangle =
    \bar{\rho}\nu_{\rm T} \nabla v_{\rm y},
    \label{eq:reynolds} 
\end{equation}
where the turbulent viscosity $\nu_{\rm T} =  u_{x}^{\prime} l$. The orange line shows the mixing length estimate from the right side of Equation~\eqref{eq:reynolds} and is again a remarkably good fit, with a mixing length $l \sim 4$ pc throughout all simulations. Since the mixing length ansatz for Reynolds stress is a good approximation, we can also construct mean shear $v_{\rm y}$ and turbulent velocity profiles $u^{\prime}$ analytically as well, though we eschew this for the sake of brevity. This suggests that the turbulent Prandtl number ($\nu_{\rm T}/D_{\rm eddy}$) is of order unity as typical in turbulent flows \citep{tennekes72}. 

\subsection{Thermal Conduction} \label{sect:3dthermalconduction}
\begin{figure}
    \centering
    \includegraphics[width=\columnwidth]{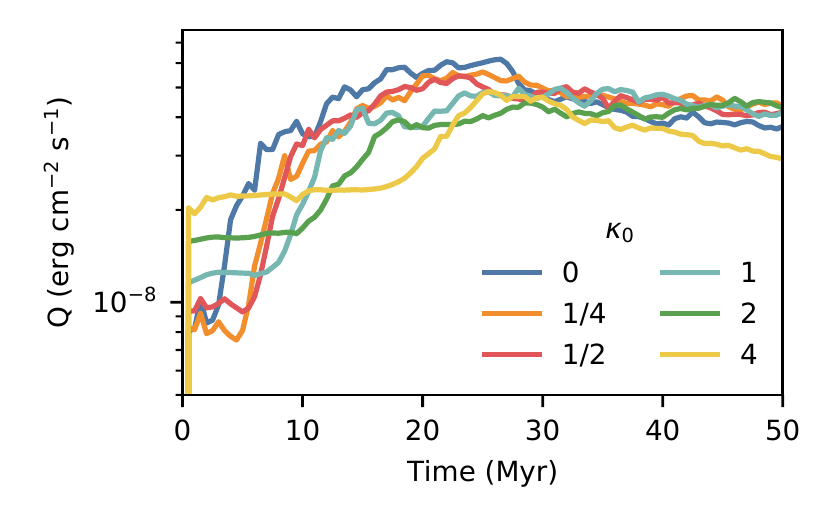}
    \caption{Dependence of cooling in the mixing layer on thermal conductivity. The curves show runs where the fiducial $\kappa$ is multiplied by a constant scaling factor $\kappa_0$. Thermal conduction does not matter until it gets large enough to suppress the turbulence.}
    \label{fig:kappa_dependence}
\end{figure}

We perform a quick assessment of the impact of isotropic thermal conduction. We defer anisotropic field-aligned conduction to future work. We use the same (constant, temperature independent) thermal conductivity as the 1D simulations, given in Eq.~\eqref{eq:conduction}, which we vary in amplitude $\kappa_0$. Note that for our fiducial case, $\kappa_{\rm turb} \sim \kappa_{\rm cond}$. The results are shown in Fig.~\ref{fig:kappa_dependence}. Conduction has no impact until $\kappa_{\rm cond} > \kappa_{\rm turb}$. At this point, $Q$ falls back towards the laminar speed $S_{\rm L}$, but $S_{\rm L}$ at this transition is already close to $S_{\rm T}$, indicating that thermal conduction is strong enough to compete with turbulent diffusion as the main source of heat transport. Without thermal conduction, we assumed that the turbulent velocity $u'$ was much larger than $S_{\rm L}$, but this assumption breaks down for strong thermal conduction since that increases $S_{\rm L}$. The scale at which $u'\sim S_{\rm L}$ is known as the Gibson scale. In turbulent combustion, below this scale, the flames burn through the turbulent eddies within an eddy lifetime and hence the speed of the front is unaffected by the turbulence. The front is thus said to be `wrinkled' by the turbulence, but not `corrugated' due to the turbulent eddies. This is also known as the `wrinkled flames regime' in the Borghi diagram.
Conduction also suppresses the multiphase nature of the mixing layer by broadening the hot/cold gas interface. Our results are consistent with previous cloud-crushing studies which show that thermal conduction hinders hydrodynamic instability but otherwise has fairly mild effects for clouds large enough to resist thermal evaporation, in conditions typical of the CGM \citep{2016ApJ...822...31B,armillotta17,Li2020MNRAS.492.1841L}. Overall, as long as turbulent diffusion dominates heat transport, conduction can be safely ignored. 

\subsection{Convergence} \label{sect:3D_convergence} 

\begin{figure}
    \centering
    \includegraphics[width=\columnwidth]{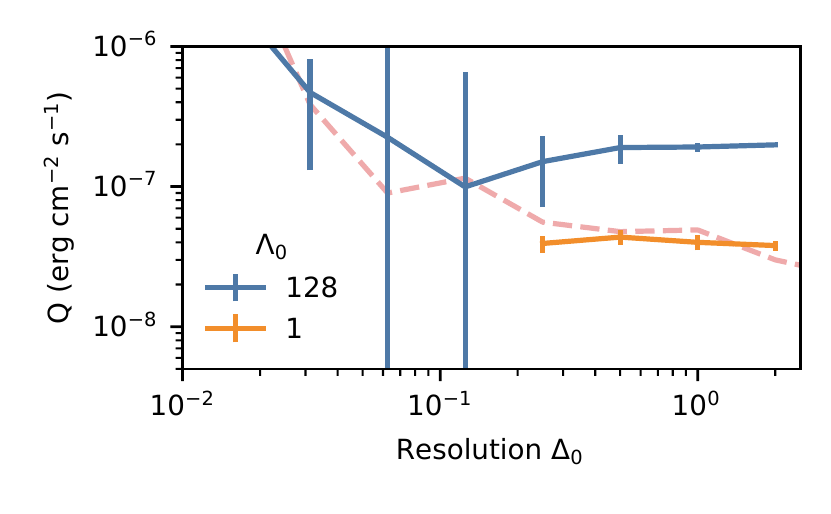}
    \includegraphics[width=\columnwidth]{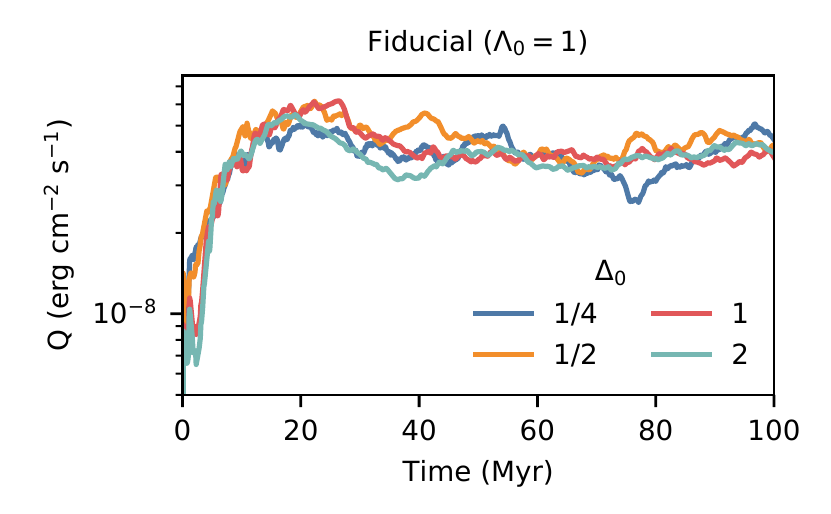}
    \includegraphics[width=\columnwidth]{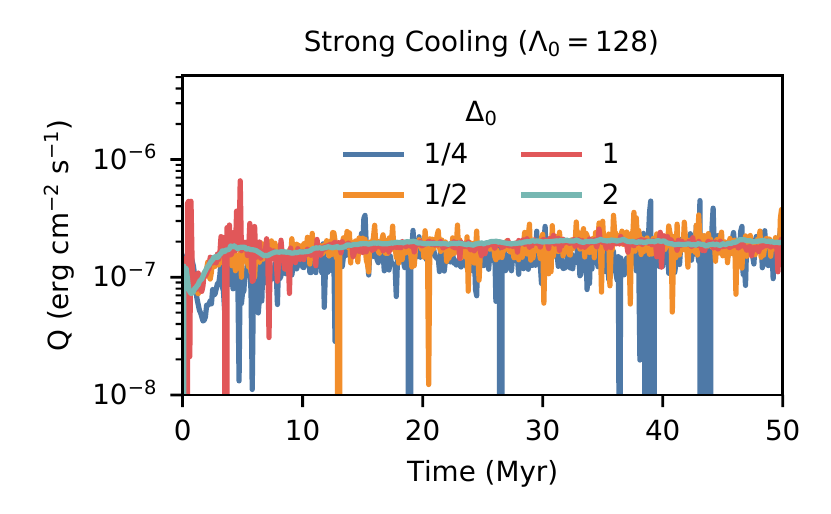}
    \caption{{\it Upper panel:} Result of varying the resolution of the 3D runs with the fiducial and strongest cooling rates. The fiducial resolution $\Delta_0 = 1$ is converged in both cases. The pink dashed line shows the result for the 1D simulations with strong cooling and no conduction for comparison (orange line in Fig.~\ref{fig:resolution}), reflecting where numerical diffusion becomes dominant. {\it Middle and Lower panels:} The time profiles of cooling for the various points shown in the top panel. The middle panel shows runs from the fiducial runs ($\Lambda_0 = 1$), while the bottom panel shows runs with the strongest cooling ($\Lambda_0 = 128$). The runs with strong cooling shows rapid oscillations that grow in amplitude as resolution is lowered.}
    \label{fig:3d_resolution}
\end{figure}

The convergence properties of this setup have been previously studied \citep{ji19,fielding20}. We therefore perform a restricted set of resolution studies for our 3D simulations, considering the fiducial case ($\Lambda_0=1$) and a case with strong cooling ($\Lambda_0=128$). The results are shown in Fig. \ref{fig:3d_resolution}. The resolutions go from a quarter to twice that of the fiducial resolution. Lower resolution runs are also shown for the run with strong cooling. Error bars are derived from fluctuations in $Q$ in the time series. In the fiducial case, we find that we are well converged, with little difference in the mean $Q$ and small error bars, indicating that the simulation is well resolved. However, for the case of strong cooling, oscillations are significantly large for lower resolutions, consistent with the 1D case, resulting in larger error bars. The cooling over time is shown in the middle and bottom panels of Fig.~\ref{fig:3d_resolution} for both cases, where we can see more clearly that in the case with fiducial cooling, the curves are generally smooth with no rapid oscillations. In the case with strong cooling, we see that as we lower the resolution, we see rapid oscillations with increasing amplitude. These oscillations have a period of roughly $t_{\rm cool}$. In 3D, the oscillations cause apparent broadening of the interface, and the cooling surface appears to adjust with resolution. However, the mean $Q$ still remain close to the converged value. As long as $Q$ is averaged over a sufficient time interval, our fiducial resolution is sufficient, even for our strongest cooling case. This is consistent with previous results of larger scale simulations, and follows the expectations from the 1D results in \S\ref{sect:1D}.

\dobib

%% file: sections/6_discussion.tex
\section{Discussion} \label{sect:disc}

\subsection{Comparison with Previous Work}

We now compare our results to recent work on radiative TMLs. We confine our comparisons to the formula for hot gas entrainment (Eq.~\eqref{eq:fiducial-scaling}) and its physical justification. 

\citet{ji19} was the first paper to confront analytic models of radiative TMLs with simulations. They pointed out that the inflow and turbulent velocities were much less than the shear velocity, that radiative cooling was balanced by enthalpy flux from the hot gas (rather than turbulent dissipation, as in e.g., \citealt{white16}), and that contrary to the widely cited model of \citealt{begelman90}, the layer width does not scale as $h \propto v_{\rm t} t_{\rm cool}$. They obtained a scaling $h\propto t_{\rm cool}^{1/2}, v_{\rm in} \propto t_{\rm cool}^{-1/2}$, which in hindsight is the scaling in the weak cooling regime; they did not run enough simulations to discern the change in slope to $v_{\rm} \propto t_{\rm cool}^{-1/4}$ in the strong cooling regime. Interestingly, they noted that using standard emissivities, mixing length theory matches temperature/density profiles well in the weak, but not strong cooling regimes. We now know this is because emissivity changes in the multiphase regime (Eq.~\eqref{eq:true_emissivity}). 

\citet{gronke19} obtained a scaling relation (Eq.~\eqref{eq:vin}) which is identical to our fiducial Eq.~\eqref{eq:fiducial-scaling} in the strong cooling regime if $u^{\prime} \sim c_{\rm s,cold}$. They consider a cold cloud embedded in a hot wind, which grows in mass and entrains. It continues to grow even after it is fully entrained ($v_{\rm shear} \rightarrow 0$); in fact, growth {\it peaks} at this point. The cloud pulsates due to loss of pressure balance seeded by radiative cooling; this in turn drives turbulence and hot gas entrainment. 

Both of these studies considered magnetic fields, which are ignored here. For instance, the plane parallel shear simulations of \citet{ji19} show that B-fields suppress turbulence and mixing due to magnetic tension, but the MHD cloud simulations of \citet{gronke19} nonetheless show cloud growth at the same rate as hydrodynamic simulations, despite very different cloud morphology in the two cases. The difference likely lies in the very different nature of turbulent driving in the shear flow and cloud scenarios, which also affects the growth in surface area. Given the substantial differences between hydrodynamic and MHD turbulence, it is important to eventually extend the arguments in this paper to the MHD case. 

Both \citet{ji19} and \citet{gronke19} invoked low pressure due to fast cooling to seed turbulence and set the entrainment rate of the hot gas, rather than the Kelvin Helmholtz instability. In \cite{ji19}, this was argued to be due to the constancy of $P + \rho u^{\prime 2} $ across the mixing layer (so that pressure drops due to cooling boost turbulence), as well as the fact that $v_{\rm in}$ appeared only weakly dependent on $v_{\rm shear}$ and independent of $\chi$, two factors which set the Kelvin Helmholtz timescale. In light of our larger suite of simulations, it is now clear that in fact entrainment rates do depend on $v_{\rm shear}$. Turbulence seeded by cooling is also inconsistent with the very weak dependence of $u^{\prime}$ with cooling strength that we see here (Fig.~\ref{fig:uprime_scalings}). For this problem, it is important to have sufficient dynamic range and dense sampling to establish scaling relations (as we have seen in the $u^{\prime}$ vs $\chi$ relation; Fig.~\ref{fig:uprime_scalings}). In this paper we argue -- consistent with results from the combustion literature, and as argued by \citet{fielding20} -- that turbulence, rather than pressure gradients, is the primary driver of hot gas entrainment. This statement has to be qualified by the fact that in the cloud case, cooling-induced pressure gradients appear to be  the primary driver of cloud pulsations and turbulence, so the end result can be the same. Thus, the \citet{gronke19} scaling for $v_{\rm in}$ still holds, as potentially do their analytic arguments\footnote[8]{E.g., they identify the timescale $t_{\rm sc,cold} \sim H/c_{\rm s,cold}$, where $H\sim (r_{\rm cl} c_{\rm s} t_{\rm cool})^{1/2}$ is analogous to the length scale in Eq.~\eqref{eq:turb_field}. This is identical to the effective cooling time (Eq.~\eqref{eq:tilde_tau}) which is critical to the model in this paper.}. However, we await detailed study of $u^{\prime}$ scalings in this scenario to refine the model. 

\citet{fielding20} ran simulations of radiative plane parallel mixing layers very similar to \citet{ji19} and this work. They rightly emphasize the fundamental role of turbulence in hot gas entrainment, and directly measure fractal properties in their simulations. They derive an analytic model whose scalings are similar to \citet{gronke19} and this work.

The analytic model of \citet{fielding20} states that: 
\begin{equation}
    v_{\rm in} = \frac{w}{t_{\rm cool}} \left( \frac{A_{\rm w}}{A_{\rm L}}  \right) = \frac{w}{t_{\rm cool}} \left( \frac{w}{L} \right)^{-1/2} ,
 \end{equation}
where $w$ is a length scale defined by $t_{\rm mix} \sim w/v_{\rm turb,hot}(w) \sim t_{\rm cool}$ and the second equality arises from fractal scalings with fractal dimension $D=2.5$, which they measure directly from their simulations. The first equality is very similar to Eq.~\eqref{eq:area_ratio}, except that  $w/t_{\rm cool}$ is substituted for $S_{\rm L}$. However, at face value, this argument would seem to imply that if the length scale $w$ is not resolved (and replaced by a resolution element $\Delta$), then inflow becomes resolution dependent, $v_{\rm in} \propto \Delta^{1/2}$. Neither \citet{fielding20} nor we see evidence for this, even in simulations where $w$ is highly under-resolved. This aspect of their model will be clarified in an upcoming paper (Fielding et al., in preparation).

\subsection{Conclusions}

Radiative mixing layers are closely analogous to turbulent combustion fronts: in both cases, the speed of front propagation $v_{\rm in}$ is determined by the temperature and density sensitive reaction rate, and thus conditions within the front itself. This is in contrast to shock propagation, where the shock speed and jump conditions are simply determined by conservation laws, independent of the small-scale details of shock structure. To obtain $v_{\rm in}$, it would seem that the structure of the front must be accurately resolved. Thus, it has long been thought that calculations of thermal fronts can only be converged if thermal conduction is included and the Field length is resolved \citep{koyama04}. Yet, recent simulations \citep{ji19,gronke18,gronke19,mandelker20a,fielding20} show remarkable robustness to resolution, despite the absence of thermal conduction -- even when the cooling front is one cell thick! They also show characteristic front propagation speeds of order the cold gas sound speed $v_{\rm in} \sim c_{\rm s,cold}$ (far less than the maximum possible $c_{\rm s,hot}$) and scalings $v_{\rm in} \propto (t_{\rm sc}/t_{\rm cool})^{-1/4}$ (where $t_{\rm sc}$ is a sound-crossing time) which are not trivial to understand. In this paper, we use models derived from the turbulent combustion literature to shed light on these issues. 

We first examine the impact of resolution on laminar fronts. The restriction to laminar fronts allows the problem to be considered in 1D, where there are analytic solutions. In the absence of thermal conduction, there is clear resolution dependence, such that $v_{\rm in} \sim \sqrt{D_{\rm num} t_{\rm cool}} \propto \Delta^{1/2}$. The numerical diffusion coefficient from truncation error is $D_{\rm num} \sim v \Delta$, $v$ is a characteristic velocity, and $\Delta$ is the grid resolution. The inclusion of conduction is indeed required for convergence. However, surprisingly it is not strictly necessary to resolve the Field length for convergence. Instead, the key requirement for convergence is that explicit thermal diffusion simply be larger than numerical diffusion: i.e., $D_{\rm cond} > D_{\rm num}$, where $D_{\rm cond} \sim \kappa/(\rho c_{\rm P})$, where $\kappa$ is the standard conduction coefficient and $c_{\rm P}$ is the specific heat at constant pressure.
If the Field length is unresolved, numerical dispersion increases, as the front structure is not accurately resolved and there are larger errors in the temperature derivatives and conductive heat flux. Nonetheless, the steady-state simulations oscillate about the correct answer. The error can be beaten down by time averaging.  This is not unusual for a stiff problem where the smallest length scale remains unresolved. 

We then examine the effects of turbulence. As in \citet{ji19}, we simulate a plane-parallel shear layer where the Kelvin-Helmholtz instability drives turbulence and mixing. We find, consistent with previous findings, that the inflow velocity $v_{\rm in}$ and surface brightness $Q$ are converged without thermal conduction. Heuristically, we argue that this is because as long as the turbulent driving scale $L$ is well resolved $L \gg \Delta$, the turbulent diffusivity $D_{\rm turb} \sim u^{\prime} L$ is always larger than the numerical diffusivity $D_{\rm num} \sim v \Delta$. Similar to our 1D results, lower resolution simply implies larger numerical dispersion and temporal oscillations in the front profile. We also find that thermal conduction has little effect unless it is larger than the turbulent diffusivity. 

The front is characterized by the dimensionless parameters:  overdensity $\chi$, Mach number ${\mathcal M}$, and most importantly the \Da number ${\rm Da}=\tau_{\rm turb}/t_{\rm cool} = L/(u^{\prime} t_{\rm cool})$, where $u^{\prime}$ is turbulent velocity at the outer scale $L$, and $t_{\rm cool}$ is the local cooling time. The \Da number, which measures the relative importance of mixing and cooling, increases as temperature falls within the front. There are two distinct regimes: 
\begin{itemize}
    \item {{\it Weak cooling} (Da < 1): the `well stirred' regime. Since the cooling time is longer than the eddy turnover time, the gas entropy is set primarily by mixing. Thus, it forms a single phase gas with smoothly varying temperature within the front. The front structure is entire analogous to a thermal conduction front, except that the conductive diffusivity $D_{\rm cond}$ is replaced by the turbulent diffusivity $D_{\rm turb}$. This implies a front thickness $h \sim (D_{\rm turb} t_{\rm cool}) \propto t_{\rm cool}^{1/2}$ and an inflow velocity $v_{\rm in} \sim (D_{\rm turb}/t_{\rm cool})^{1/2} \propto t_{\rm cool}^{-1/2}$}. 
    \item{{\it Strong cooling} (Da > 1): the `corrugated flamelet' regime. In this limit, the cooling time is shorter than the mixing time, the gas entropy is set primarily by cooling, and the gas fragments into a multiphase medium. The interface between the two phases is highly corrugated, and has been shown to have a fractal structure \citep{fielding20}. This increase in surface area of the front boosts the mass flux across the front.
The surface area increase is resolution dependent. Nonetheless, hot gas as a whole is consumed at a rate $v_{\rm in} \sim u^{\prime}$ (where $u^{\prime}$ is the turbulent velocity at the outer scale), independent of resolution.
The rate limiting step in determining hot gas consumption is the turbulent mixing rate, which proceeds at the outer scale velocity $u^{\prime}$. It is similar to how mixing time of a passive scalar (e.g., cream in coffee) is set by the eddy turnover time at the outer scale, independent of the details of molecular diffusivity. Rapid mixing in both cases depends on the large increase in surface area due to turbulence.} 
\end{itemize}

Our results are also of importance to the resolution requirements in larger scale simulations; e.g., cosmological simulations which are currently unconverged in the cold gas properties \citep{2016MNRAS.461L..32F, hummels19}. Ultimately, the physics of radiative TMLs explored here sets the mass and momentum transfer between the hot and the cold phase, and thus, impacts not only the morphology of the multiphase medium but also, for instance, the fuel supply for future star-formation. 
In this work, we showed that as long as numerical diffusion is not the dominant mixing mechanism, it is sufficient in the presence of turbulence to resolve the outer scale $L$ of the turbulence to obtain a converged solution, and not the width of the laminar front, contrary to conventional wisdom. While in many astrophysical applications $L$ is likely of the order of $\sim $parsecs \citep{mccourt18,gronke18} and thus challenging to resolve directly in large scale simulations, our findings relax the resolution requirements by up to orders of magnitude.

At a more detailed level, one must still take into account the behavior of small scales in the strong cooling regime. We argue that there is a characteristic effective cooling timescale $\tilde{\tau}_{\rm cool} \sim \sqrt{(L/u^{\prime}) t_{\rm cool}}$. This effective cooling time is resolution independent. The turbulent velocity $u^{\prime}$ measured in the simulations peaks in the multiphase region where the cold gas fraction is $\sim 50\%$, where cooling also peaks. A similar lifetime for eddies in combustion fronts was given by \citet{gulder91}, by assuming that the mixing front is corrugated down to the Taylor microscale.  Thus, the front propagates at a velocity $v_{\rm in} \sim (D_{\rm turb}/\tilde{\tau}_{\rm cool})^{1/2} \propto u^{\prime 3/4} t_{\rm cool}^{-1/4}$. Our fiducial scalings are given by Eqs.~\eqref{eqn:Q_slow} and \eqref{eqn:Q_fast} in the slow and fast cooling regimes respectively.
The slow cooling result is a straightforward application of 1D mixing length theory, while the fast cooling result agrees well with previous simulation work (Eq.~\eqref{eq:vin}) if $u^{\prime} \sim c_{\rm s,cold}$ (see below).  

We have verified directly in our simulations the scalings $v_{\rm in}, Q \propto u^{\prime 1/2} t_{\rm cool}^{-1/2}$ and $v_{\rm in}, Q \propto u^{\prime 3/4} t_{\rm cool}^{-1/4}$ in the weak and strong cooling regimes respectively (Figs.~\ref{fig:coolvscool} and \ref{fig:shear_dependence}). We also show that are no hidden parameters; and in particular no dependence on overdensity $\chi$ or Mach number. Astrophysically, the strong cooling regime is often of more interest. For instance, for clouds embedded in a hot wind to survive cloud-crushing instabilities, $t_{\rm cool}(T_{\rm mix}) < t_{\rm cc} \sim L/u^{\prime}$ \citep{gronke18}, where $T_{\rm mix} \sim (T_{\rm hot} T_{\rm cold})^{1/2}$, which implies that most emission is in the strong cooling regime. In this regime, we verified in our simulations that within the front, the cooling rate tracks the cold gas fraction (which tracks the surface area), peaking at $f_{\rm cold} \sim 50\%$, and that the cooling flux has a Gaussian shape (Fig. \ref{fig:phase_profiles}), as expected for the front position if it undergoes a random walk. In addition, we show that the effective emissivity in the multiphase region of the simulations scales as $\tilde{\epsilon} \sim P/\tilde{\tau}_{\rm cool} \propto \sqrt{u^{\prime} t_{\rm cool}}$, in agreement with our model. The emissivity $\tilde{\epsilon} \propto u^{\prime 1/2}$ tracks turbulence and hence the shearing rate. The width of the strong cooling region also obeys an analytic scaling relation Eq.~\eqref{eq:h_fast_cooling} predicted by theory. If we use a turbulent diffusion coefficient and emissivity $\epsilon,\tilde{\epsilon}$ in the weak (strong) cooling regimes respectively, we can match temperature and density profiles in the simulations with mixing length theory remarkably well. The turbulent velocity follows mixing length scalings $u^{\prime} \approx l \nabla v_{\rm y}$, and the Reynolds stress is also accurately represented by mixing length theory (Fig.~\ref{fig:Reynolds_stress}). This allows for rapid construction of semi-analytic profiles of radiative mixing layers without recourse to simulations, which is very useful for comparing against observations. 

All that remains is to specify the turbulent velocity at the outer scale, $u^{\prime}$. Turbulence can arise either from external driving (in this paper, due to the shear flow), or be driven by radiative cooling itself (e.g., clouds with $r > c_{\rm s} t_{\rm cool}$ which lose sonic contact with their surroundings and begin to pulsate;  \citealt{gronke19,gronke20}). In this shear driven case, we have verified that $u^{\prime} \propto \nabla v_{\rm shear}$, as predicted by mixing length theory. To order of magnitude, $u^{\prime} \sim c_{\rm s,cold}$ for cloud pulsations or transonic ${\mathcal M} \sim 1$ shear flows, but here we find detailed parameter dependences. For the plane-parallel shear flow in these simulations, we find that $u^{\prime} \propto v_{\rm shear}^{0.8}$, almost no dependence on cooling time $t_{\rm cool}$, and a non-monotonic dependence on overdensity (Fig. \ref{fig:uprime_scalings}). For $\chi \gtrsim 100$, Eq.~\eqref{eq:uprime_scalings} is a reasonable approximation. These scalings will of course depend on the nature of turbulent driving. In the future, we plan to investigate turbulent scalings in pulsating clouds, and the effect of B-fields on these scalings. We stress, however, that Eq.~\eqref{eq:fiducial-scaling} is general, independent of the source of turbulent driving.  

In summary, the cold gas mass growth rates we find in our 3D simulations agree with our analytic model (\S\ref{sect:fiducial_scaling}) and read:
\begin{equation}
    v_{\rm in}\approx 
    11.3 {\mathrm{~km}\,\mathrm{s}^{-1}} 
    \mathcal{M}_{\rm turb}^{1/2}
    \left(\frac{L}{100\,{\rm pc}}\right)^{1/2} 
    \left(\frac{t_{\rm cool,cold}}{0.03\,{\rm Myr}}\right)^{-1/2}
\end{equation}
for the ${\rm Da} < 1$ `well stirred' (slow cooling) regime, and 
\begin{equation}
    v_{\rm in}\approx 
    9.5 {\mathrm{~km}\,\mathrm{s}^{-1}}
    \mathcal{M}_{\rm turb}^{3/4} 
    \left(\frac{L}{100\,{\rm pc}}\right)^{1/4}
    \left(\frac{t_{\rm cool,cold}}{0.03\,{\rm Myr}}\right)^{-1/4}
\end{equation}
for the ${\rm Da} > 1$ `corrugated flame' (fast cooling) regime. Here, $\mathcal{M}_{\rm turb}\equiv u' / c_{\rm s, cold}$ but as stated above $u'$ -- unlike the $v_{\rm in}$ scalings -- depends on the geometry employed. For shearing layers, we find (\S\ref{sect:uprime_scalings}) that
\begin{equation}
    u' \approx 50{\mathrm{~km}\,\mathrm{s}^{-1}}
    \mathcal{M}^{4/5}
    \left(\frac{c_{\rm s,c}}{15 {\mathrm{~km}\,\mathrm{s}^{-1}}} \right)^{4/5}
    \left(\frac{t_{\rm cool,cold}}{0.03\,{\rm Myr}}\right)^{-0.1} ,
\end{equation}
for $\chi\gtrsim 100$ and $\mathcal{M}\equiv v_{\rm shear} / c_{\rm s,hot}$ as used throughout.

Of course, at higher levels of precision, details of the interaction between turbulence, diffusion and cooling remain to be explored. Just as there are a plethora of models and computational algorithms for subgrid turbulent scalar transport (often used in simulations of metal mixing), there are a plethora of models for subgrid turbulent combustion (e.g., see \citealt{swaminathan11}). The issues are more complex since combustion can backreact on the flow and change its properties. Such models have been used in simulations of thermo-nuclear burning on white dwarfs \citep{schmidt06,jackson14}, where the burning fronts are unresolved. Such sub-grid models would be an interesting avenue for future work, particularly in the context of cosmological simulations of galaxy formation, where the separation of scales is even more forbidding than in Type Ia supernova problem. Another avenue for more detailed future work is the inclusion of non-equilibrium chemistry. In this work, equilibrium cooling curves were used in all simulations. In reality, material will often be out of equilibrium, with recombination/ionization rates often having time scales longer than the cooling time. Metallicity differences between the different phases could also lead to further complications beyond the analysis in this paper. While non-equilibrium ionization/recombination was taken into account in \citep{kwak10,ji19}, to our knowledge there have not been any studies which incorporate non-equilibrium cooling. Such details could be important due to their bearing on the predictions of observables such as column densities. It is also important to continue to verify scalings for $S_{\rm T}$ in higher resolution simulations, perhaps with thermal conduction so that $S_{\rm L}$ is well-defined and resolution-independent. One important limitation of current simulations is that the front width $\delta$ and Kolmogorov scale $\eta$ are unresolved and simply equal to the grid scale, so that Ka~$\sim (\delta/\eta)^{2} \sim 1$. In practice, these scales could be sufficiently separated (with Ka~$\gg 1$) that the arguments in \S\ref{sect:fiducial_scaling} no longer apply. If so, turbulence can penetrate the conductive interface and affect conditions there, impacting the effective and total cooling rates. An intriguing approach in the spirit of 1D modeling in this paper, and useful for developing physical insight is the Linear Eddy Model \citep{kerstein88}, which attempts to model the effects of turbulence in 1D so that extremely high resolution can be achieved, and has good support from experiments and direct numerical simulation.  It has been successfully applied to the Type Ia supernova problem \citep{woosley09}. These are promising avenues for future work. 

\dobib

%% file: sections/7_ackd.tex
\section{Acknowledgements}
We thank Omer Blaes, Drummond Fielding, Yan-Fei Jiang, Chris White, and Nir Mandelker for helpful discussions, and the referee Evan Scannapieco for a helpful and constructive report. We also thank Yan-Fei Jiang for the use of his conduction module for Athena\verb!++!, and Suoqing Ji for providing his setup in FLASH. We have made extensive use of the yt astrophysics analysis software suite \citep{yt}, matplotlib \citep{Hunter2007}, numpy \citep{van2011numpy}, and scipy \citep{2020SciPy-NMeth} whose communities we thank for continued development and support. We acknowledge support from NASA grants NNX17AK58G, 19-ATP19-0205, HST theory grants HST-AR-15039.003-A, HST-AR-15797.001-A, XSEDE grant TG-AST180036, and the NASA Hubble Fellowship grant HST-HF2-51409. 

\section{Data Availability}
The data underlying this article will be shared on reasonable request
to the corresponding author.

\dobib

%% file: sections/8_appendix.tex
\appendix
\section{Code Verification Test: 1D Diffusion Couple} \label{appendix:conduction} 
\begin{figure}
    \centering
    \includegraphics[width=\columnwidth]{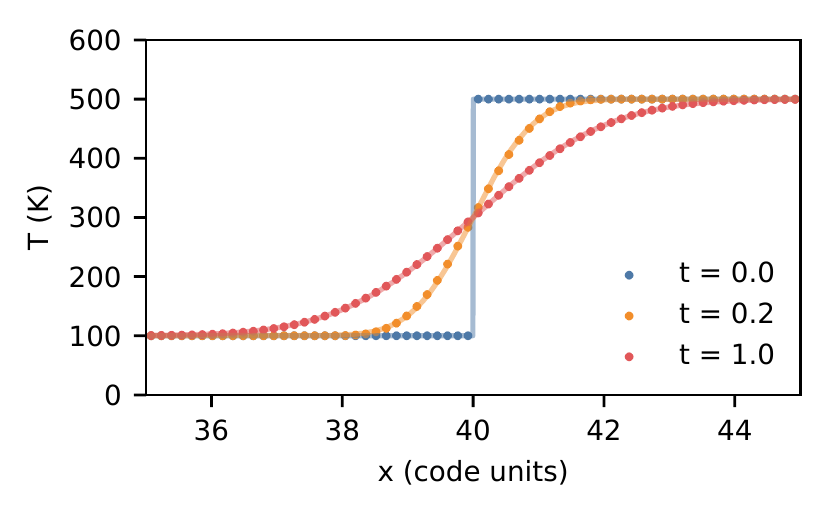}
    \caption{Results of the implementation test for thermal conduction. The data points          from simulation at various times are shown to match the analytic solution,          given by the solid lines.}
    \label{fig:tc}
\end{figure}
We test the conduction module by considering the interface between two regions of different temperatures that is initially represented by a step function located at $x=x_0$ when $t=0$. The left side is at temperature $T_1$, and the right side is at $T_2$. The analytical solution is then given by:
\begin{equation}
    T(x,t) = T_0 + \frac{\Delta T}{2}\erf(\frac{x - x_0}{2\sqrt{\alpha t}}),
\end{equation}
where $T_0$ is the mean of $T_1$ and $T_2$, and $\Delta T = T_2 - T_1$. This solution assumes that density is fixed, so we turn off the hydrodynamics updates to the density field and velocity fields, and only let the energy of the simulation cells evolve.

\begin{table}
    \centering
    \begin{tabular}{ |c|c|c|c|c|c|c|c| } 
     \hline
     Resolution & $T_1$ & $T_2$ & $x_0$ & $\gamma$ & $\rho$ & $\kappa$ & $V_m$ \\ 
     \hline
     512 & 100 & 500 & 40 & 5/3 & 0.75 & 1.5 & 10 \\ 
     \hline
    \end{tabular}
    \caption{Parameters used for the thermal conduction test.}
    \label{table:cdn_tbl}
\end{table}

We choose the set of parameters listed in Table~\ref{table:cdn_tbl}, and ensure that the chosen value of $V_m$ is sufficiently high for a well converged solution. The results are shown in Fig.~\ref{fig:tc}, which show that the simulation data is a good match to the analytical solution. The code is also verified for a case where the density is not held constant in the resolution tests for 1D thermal fronts described in \S\ref{sect:1D}, where the integrated cooling over a steady thermal front is shown to converge to the expected analytical result.